\documentclass{article}
\usepackage[a4paper]{geometry}
\usepackage{amsmath}
\usepackage{graphicx}
\usepackage{color}
\usepackage{parskip}
\usepackage{subfigure}

\usepackage{multibib}
\def\degree{{k}}
\def\linksRC{k^+}

\def\richClubRank{{\Phi}}
\def\richClubDegree{{\phi}}
\def\weight{{w}}
\def\weightPlus{{w^+}}

\def\eProb{p}
\def\averageDegDeg{k_{{\rm nn}}}
\def\entropy{S}

\def\stepFunct{{s}}

\def\walk{W}
\def\jumpProb{P_{i\rightarrow j}}
\def\firstEigenvalue{\Lambda_1}
\def\firstEigenvector{\overline{v}}
\def\degreeCF{k_{\rm cut}}
\def\degreeMax{k_{\rm max}}
\def\eigenval{\Lambda}

\title{Beyond the rich-club: Properties of networks related to the better connected nodes.}
\author{R. J. Mondrag\'on\\[2ex]
School of Electronic Engineering and Computer Science\\ Queen Mary University of London\\
Mile End Rd., London E1 4NS, UK\\
{\tt r.j.mondragon@qmul.ac.uk}}

\begin{document}
\maketitle

\section*{Abstract}
Many of the structural characteristics of a network depend on the connectivity with and within the hubs.  These  dependences can be related to the degree of a node and the number of links that a node shares with nodes of higher degree. In here we revise and present new results showing how to construct network ensembles which give a good approximation to the degree--degree correlations, and hence to the projections of this correlation like the assortativity coefficient or the average neighbours degree. We present a new bound for the structural cut--off degree based on the connectivity within the hubs. Also we show that the connections with and within the hubs can be used to define different networks cores. Two of these cores are related to the spectral properties and walks of length one and two which contain at least on hub node, and they are related to the eigenvector centrality. We introduce a new centrality measured based on the connectivity with the hubs. In addition, as the ensembles and cores are related by the connectivity of the hubs, we show several examples how changes in the hubs linkage effects the degree--degree correlations and core properties. 

\section{Introduction}
What does the Internet, the human brain and the super--heroes have in common? If the connectivity of the Internet, the brain and the friendship between the super--heroes is represented with a network, there exist a small set of nodes which have a large numbers of links, the so--called rich nodes, hubs or stars~\cite{Zhou04,Heuvel11,gleiser2007become}. Rich nodes can or cannot have connections between themselves. If they do, we can interpret this as the presence of a core, a set of well connected nodes that are well connected between themselves. 

The connectivity within rich nodes has  been associated with many structural characteristics of a network like assortativity~\cite{Xu10}, clustering coefficient~\cite{Xu10}, existence of motifs~\cite{xu2011changing}, the stability of dynamical processes~\cite{gollo2015dwelling} and the construction of network's null--models~\cite{Mondragon2014}. 
The aim here is to bring some of these previous and new results together by showing the dependance of these network properties with the node's degree and the number of links that a node shares with nodes of higher degree. 

\begin{figure}
\begin{center}
\includegraphics[width=10cm]{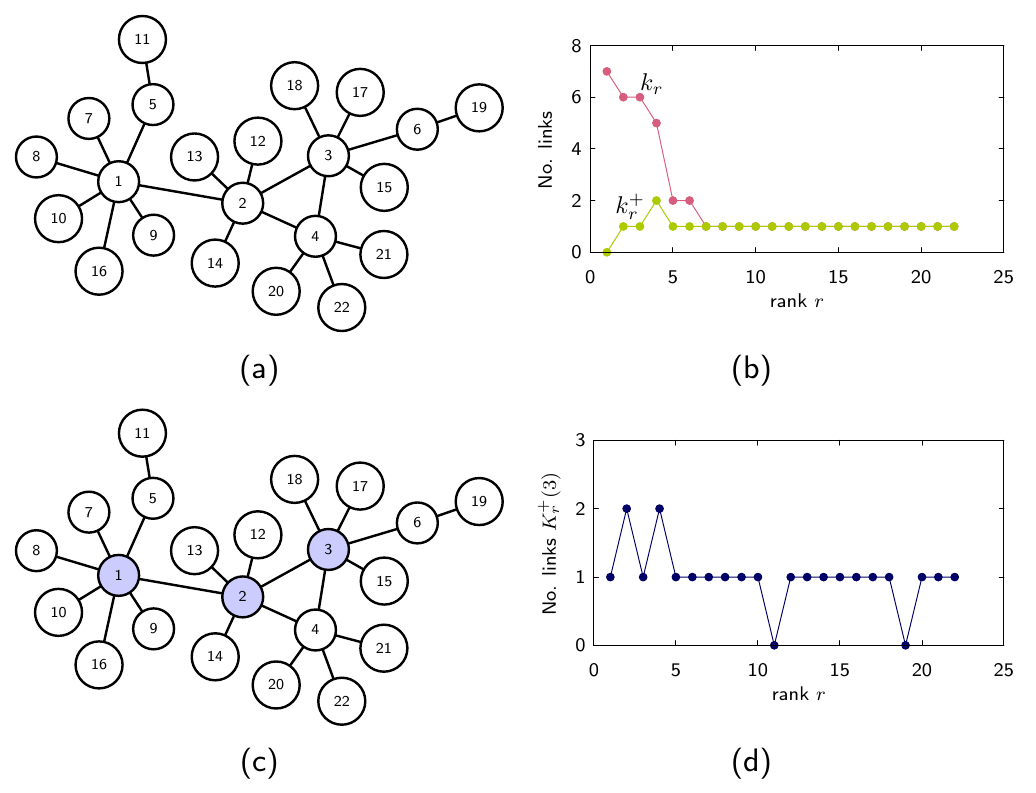}
\end{center}
\caption{\label{fig:uno} 
(a) A network where its nodes are ranked in decreasing order of their degree and (b) the degree $k_r$ of node $r$ and  the number of links $\linksRC_r$ that node $r$ has with nodes of higher rank. (c) The same network where the three top ranked nodes form a core and (d) number of links $K^+_r(3)$ that node $r$ has with the any of these three top ranked nodes.
}
\end{figure}

In an undirected network the connectivity of its nodes is described by their degree $\degree$. Two of the simplest properties of a network are its maximum degree $\degreeMax=\max(\degree_i)$, $i=1, \ldots, N$ and its average degree $\langle \degree \rangle =\sum_{i=1}^N\degree_i/N=2L/N$, where $N$ is the total number of nodes and $L$ is the total number of links.
A first step to describe a network is via the degree sequence $\{\degree_i\}$, $i=1,\ldots,N$ which it is used to measure the network's degree distribution $P(\degree)$, that is, the fraction of nodes with degree $\degree$. The degree distribution gives only partial information about the network structure, a better description can be obtained from the number of links that a node shares with nodes of higher degree. Here, we assume that the sequence $\{\degree_i\}$ contains the node's degree ranked in decreasing order, that is, node 1 has the largest degree $k_1$, node 2 the second largest degree $k_2$ and so on. The sequence $\{\linksRC_i\}$ describes the number of links that node $i$ has with the $i-1$ nodes of higher rank (see Fig.~\ref{fig:uno}(a)-(b)). The term $\linksRC_i$ is bounded by the degree $\linksRC_i \le \degree_i$ and satisfies that $\sum_{i=1}^N \linksRC_i= L$. For networks where multilinks are not allowed, $\linksRC_i$ satisfies the bound $\linksRC_i \le i-1$, that is, the node of rank $i$ cannot have more than $i-1$ links with the $i-1$ nodes of larger rank. 

The other sequence considered here is defined by first taking a subset of the top $r$ ranked nodes, then $K^+_i(r)$ describes the number of links that node $i$ has with this set (see Fig.~\ref{fig:uno}(c)-(d)). The sequence $\{K_i^+(r)\}$ gives information about the ``influence'' that the top $r$ nodes have on the network.  The sequences $\{\degree_i$\}, $\{\linksRC_i\}$ and $\{K_i^+\}$ can be  extended to weighted networks. If the nodes are ranked in decreasing order of their weight $w_i$, then $w_i^+$ would be the total weight that node $i$ shares with nodes of greater rank and $W_i^+(r)$ would be the total weight that node $i$ shares with the top $r$ nodes.

In section~\ref{sec:Correlations} we revise and extend some previous results showing that from the sequences $\{\degree_i\}$ and $\{\linksRC_i\}$ it is possible to build network ensembles where the degree--degree correlation is well defined from the data. We introduce a new structural cut--off degree in the case that the ensemble describes the average connectivity of the networks. This section also contains several examples to show the association between the connectivity of the hubs and assortativitiy and clustering coefficients.

In section~\ref{sec:cores}, again we revise some previous results showing how the sequences $\{\degree_i\}$, $\{\linksRC_i\}$ and $\{K_i^+\}$ can be used to define different cores of the network, including the rich--club. We show that some cores based on the connectivity of the well connected are closely related to the eigenvector centrality and how can be used to define a network's core and a new centrality measure based on a core--biased random walker. The section ends with comments about the communicability~\cite{estrada2008communicability} and the time evolution of the cores.

We end with a Discussion section and there are two appendices at the end containing supplementary information.

\section{\label{sec:Correlations} Ensembles and correlations}

As mentioned before, the first step to describe a network is via its degree sequence $\{\degree_i\}$.  A better description can be obtained from the degree--degree correlation $P(\degree,\degree ')$, the probability that an arbitrary link connects a node of degree $\degree$ with a node of degree $\degree '$.  In scale--free networks  it is not possible from network's measurements to evaluate accurately the degree--degree correlation due to the small number of nodes with high degree and the finite size of the network, hence, the structure of the network is characterised using different projections of the degree--degree correlation, like the assortativity coefficient $\rho$~\cite{Newman02} or the average degree of the nearest neighbours $\langle \averageDegDeg(k)\rangle$~\cite{pastor2001dynamical}.

A network with positive assortativity coefficient has the property that nodes of high degree tend to connect to nodes of high degree, this property would be reflected in the sequence $\{\linksRC_i\}$ as it is expected that the well connected nodes share connections with other well connected nodes, then for these nodes, $\linksRC_i$ has a relative large value. 

\subsection*{Maximal entropy approach}

The information contained in $\{\degree_i\}$ and $\{\linksRC_i\}$  can be used to construct a network ensemble via Shannon's entropy~\cite{bianconi2007entropy,bianconi2009entropy,squartini2011analytical,johnson2010entropic,hou2014maximum,squartini2015unbiased}.
The Shannon entropy for a network is $\entropy = -\sum_{i=1}^N \sum_{j=1; j\ne i}^N \eProb_{ij} \log( \eProb_{ij} )$ where $\eProb_{ij}$ is the probability that $i$ shares links with node $j$. The maximisation of this entropy is attractive because it produces null-models with probabilistic characteristics only warranted by the data. 
The ensemble obtained from the maximisation of the entropy satisfy the {soft constraints} $\langle \linksRC_r \rangle = L\sum_{i=1}^r\eProb_{ir}=\linksRC_r$ and
$\langle \degree_r \rangle = L\sum_{i=1}^N\eProb_{ir}=\degree_r$, where  the total number of links $L$ is conserved and no self--loops are allowed, i.e. $p_{rr}=0$.
The maximal entropy solution under these constraints is given by the probabilities~\cite{Mondragon2014} \begin{equation}
\label{eq:mainResult}
\eProb_{ij} = 
\frac{\stepFunct(i)\left(\degree_i-\linksRC_i \right) }{\sum_{n=1}^{j-1} \stepFunct(n)\left(\degree_n-\linksRC_n \right)}\frac{ \linksRC_j}{L},\quad i <j 
\end{equation}
where 
\begin{equation}
\label{eq:stepFunct}
\stepFunct(m) = \frac{\stepFunct(m-1)\sum_{i=1}^{m-1}\stepFunct(i)(\degree_i-\linksRC_i)}{\sum_{i=1}^{m-1}\stepFunct(i)(\degree_i-\linksRC_i)-\linksRC_m\stepFunct(m-1)}.
\end{equation}
The values of $\stepFunct(m)$  are defined recursively with the initial condition $\stepFunct(1)=1$.  As we are considering undirected networks $\eProb_{ij}=\eProb_{ji}$. 
The average number of links between nodes $i$ and $j$ is $e_{ij}=Lp_{ij}$ with variance ${\rm var}(e_{ij})=Lp_{ij}(1-p_{ij})$. 
This maximal entropy solution can be used to construct the following ensembles: 
\begin{itemize}
\item[ME1] If the sequences $\{ k_i\}$ and $\{\linksRC_i\}$ are conserved, 
the networks from the ensemble have similar correlations as the original network. This ensemble has been studied before~\cite{Mondragon2014} but here we extend it as follows.
\item[ME2] If the sequence $\{ k_i\}$  is given but the sequence $\{\linksRC_i\}$ is defined up to the constraint $\linksRC_i \le r-1$, then the ensemble would have the same degree sequence and on average two nodes would have only one link. 
\item[ME3] If in the ME2 ensemble we remove the restriction $\linksRC_i \le r-1$, then the ensemble would have the same degree sequence but it is possible to have, on average, more than one link per pair of nodes.
\end{itemize}
These ensembles produce networks with different statistical properties which can be measured via the average neighbour degree or the assortativity coefficient.
The first ensemble consist of networks with similar correlation than the data. The second ensemble consist of networks where the correlation is zero if the maximal degree $\degreeMax$ is smaller than the structural cut--off degre $\degreeCF=\sqrt{N\langle k\rangle}$.
 If the the maximal degree is greater than cut--off degree then the network is correlated due to structural constraints~\cite{pastor2001dynamical} and it is not possible to construct an uncorrelated network without introducing multiple links between nodes.
 The third ensemble produces uncorrelated networks if multiple links between nodes are allowed.  It is not difficult to generate a network that is a member of one of the previous ensembles, the network is generated using a Bernoulli process where the existence of a link between nodes $i$ and $j$ is given by $\eProb_{ij}$. 

The average nearest neighbours degree given by $\langle k_{\text{nn}}(k) \rangle =\sum_{k'} k' P(k'|k)$,
where $P(k'|k)$ is the conditional probability that given a node with degree $k$ its neighbour has degree $k'$. For an uncorrelated network $\langle \averageDegDeg(k) \rangle = {\langle k^2 \rangle}/{\langle k \rangle}$. In our case~\cite{Mondragon2014}
\begin{equation}
\langle k_{\text{nn}}(k)\rangle = \frac{1}{N_k}\sum_{i=1}^{N}\left(\frac{1}{k}
\sum_{j=1}^{N} \eProb_{ij}L k_j\right)\delta_{k_i,k},
\label{eq:knn}
\end{equation}
where $\delta_{k_i,k}=1$ if $k_i=k$ and zero otherwise.
The assortativity coefficient is given by
\begin{equation}
\label{eq:assoCoef}
\rho =\frac{\left\langle k k' \right\rangle_\ell- \left\langle k \right\rangle^2_\ell}{\left\langle k^2\right\rangle_\ell-\left\langle k  \right\rangle_\ell^2}
=\frac{\sum_{i=1}^N\sum_{j=1}^N p_{ij}k_i k_j-\left( \sum_{i=1}^N\sum_{j=1}^N p_{ij}(\degree_i+\degree_j)/2\right)^2}{\sum_{i=1}^N\sum_{j=1}^N p_{ij}(\degree_j^2+\degree_i^2)/2-\left( \sum_{i=1}^N\sum_{j=1}^N p_{ij}(\degree_i+\degree_j)/2\right)^2}
\end{equation}
where $\langle \ldots\rangle_\ell$ is the average over all links. 

\begin{figure}
\begin{center}
\includegraphics[width = 10cm]{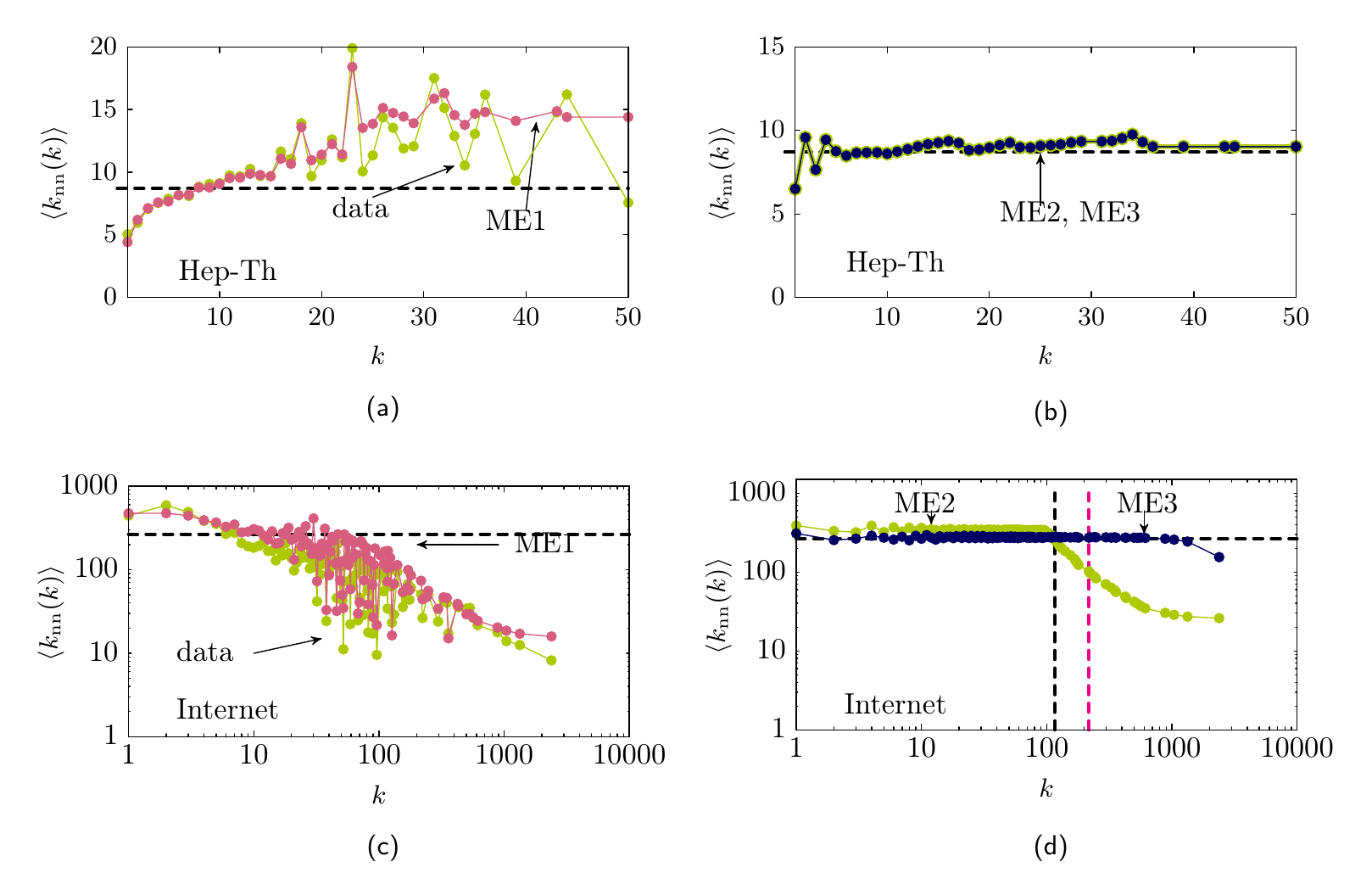}
\end{center}
\caption{\label{fig:corrEnsembles} (a) Average neighbours degree $\langle \averageDegDeg(k) \rangle$ for the Hep-Th network and the ensemble ME1.
(b) For the Hep-Th the ensembles obtained by the ME2 and ME3 methods produce uncorrelated networks. (c) Average neighbours degree for the AS-Internet network and the ensemble obtained using the  ME1 method showing that ME1 approximates well $\langle \averageDegDeg(k)\rangle$. (d) For the AS-Internet the ME2 and ME3 produce ensembles that show correlations for the links that include nodes of high degree. The orange dotted line marks the value of the cut-off degree $\degreeCF=\sqrt{N\langle k \rangle }$. The grey dotted vertical line shows the structural cut--off degree obtained from Eq.~(\ref{eq:myCutoff}).  The dashed horizontal line shows the value $\langle \averageDegDeg(k)\rangle=\langle k^2 \rangle /\langle k\rangle$ for the uncorrelated network.
}
\end{figure}

As an example, Fig.~\ref{fig:corrEnsembles} show the average neighbours degree for the Hep-Th network and the AS-Internet network. For both networks the ME1 method produces ensembles with similar correlations as the  original network (see Fig.~\ref{fig:corrEnsembles}(a) and \ref{fig:corrEnsembles}(c)). For the Hep--Th network its maximal degree is less than the cut--off degree so it is expected that the maximal entropy networks produced by the ME2 and ME3 methods generate uncorrelated networks (Fig.~\ref{fig:corrEnsembles}(b)).
For the AS--Internet the maximal degree is greater than the structural cut--off degree $\degreeCF$, in this case the ME2 
produce ensembles where only the links with end nodes of degree lower than $\degreeCF$ are uncorrelated (Fig.~\ref{fig:corrEnsembles}(d)). For the ME3 ensemble, multilinks are allowed, and the correlation shown in the figure is due to the structural constraint that self--loops are no allowed.

In the case of weighted networks, Eqs.~(\ref{eq:mainResult})-(\ref{eq:stepFunct}) are still valid.  In this case the links have weights $w_i$ and the network is described by the sequences $\{w_i\}$ and $\{w_i^+\}$ instead of $\{\degree_i\}$ and $\{\linksRC_i\}$.
For the ME3 ensemble the probabilities $\eProb_{ij}$ are well approximated via the configuration model $p_{ij}=(\weight_i\weight_j)/L^2$, and $\langle w^+_i \rangle$ is well approximated via 
\begin{equation}
\label{eq:approConfModel}
\langle w_i^+ \rangle = L\sum_{n=1}^{i-1}\frac{\weight_i\weight_n}{L^2} = \frac{\weight_i}{L}\sum_{n=1}^{i-1}\weight_n.
\end{equation}
From this equation we can evaluate the structural cut--off degree for unweighted networks. If we consider that $\linksRC_i$ is approximated via $w^+_i$ and in networks where multilinks are not allowed $\linksRC_i \le i-1$ then the condition for a multilink is
\begin{equation}
\label{eq:myCutoff}
\linksRC_i\approx \frac{\weight_i}{L}\sum_{n=1}^{i-1}\weight_n > i-1.
\end{equation}
The structural cut--off degree corresponds to the node with the largest rank $i$ where the above condition is true (see Fig.~\ref{fig:corrEnsembles}(d)). We can consider this structural cut--off also as a core of the network, these are the nodes that due to the structural constraints introduce correlations between the nodes.

\subsection*{Restricted randomisation}
The maximal entropy approach generates (canonical) ensembles with the soft constraints $\langle \degree_i\rangle = \degree_i$ and $\langle \linksRC_i \rangle = \linksRC_i$. In the case that what it is required is an ensemble where the networks ensembles satisfies hard constraints (micro--canonical), that is that the sequences $\{\degree_i\}$ and $\{\linksRC_i\}$  are conserved, the approach is to generate the ensemble numerically using a restricted randomisation approach~\cite{maslov2002specificity}. As in the case of the ME1 ensemble, the networks obtained from the restricted randomisation have similar degree--degree correlations as the original network~\cite{Mondragon2012}. For the case that maximal degree is smaller than the structural cut--off degree and only the degree sequence is conserved, the restricted randomisation would generate  uncorrelated networks as the ensembles ME2 and ME3, in this case the link probabilities are well approximated by the configuration model. If the maximal degree is larger than the cut--off degree then the networks forming the restricted randomisation ensemble would have different degree--degree correlations than the networks from  the ME2 and the ME3 ensembles~\cite{squartini2015unbiased,squartini2015breaking}.

\subsection*{Clustering and correlations}
In networks where the structure can be fully described from the degree distribution and the degree--degree correlation, the expected number of triangles and hence the clustering coefficient can be evaluated from the ensemble~\cite{dorogovtsev2010lectures}.

\begin{figure}
\begin{center}
\includegraphics[width=10cm]{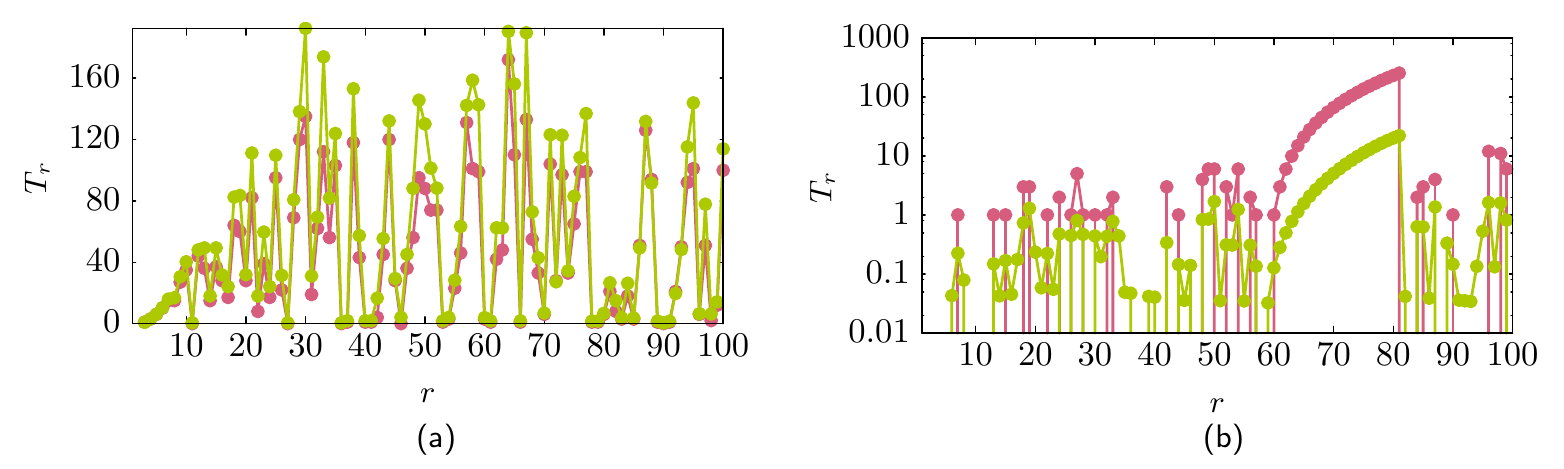}
\end{center}
\caption{\label{fig:clusterInter} Number of triangles $T_r$ and its approximation using the ME1 ensemble (a)  for the AS--Internet and
(b) for the Hep--Th. }
\end{figure}
Fig.~\ref{fig:clusterInter}(a)--(b) show the number of triangles $T_i$ that node $i$ has with nodes $j$ and $k$ which are of higher rank, $i>j>k$ and the average number of triangles $\langle T_i \rangle$ obtained from the ME1 ensemble.  Notice that for this ensemble, due to the soft constraints, it is possible to have more than one link between two nodes and this could have a large effect on the number of triangles.  Fig.~\ref{fig:clusterInter}(a) shows the results for the AS--Internet where the approximation $T_i$ via $\langle T_i \rangle$ is good because the structure of the 1997 AS--Internet can be described with the degree distribution and the degree--degree correlation~\cite{bianconi2005loops,bianconi2006effect,Zhou04}.
Fig.~\ref{fig:clusterInter}(b) shows the case for the Hep--Th network. In this case the number of triangles of the network differs considerably from the ME1 ensemble because to fully describe the structure of this network we need higher order correlations. However, even that $\langle T_i \rangle$ can be orders of magnitude less than  $T_i$, the trend between these two quantities is similar, verifying that the degree correlations strongly affect the frequency of triangles~\cite{bianconi2006effect}.

\begin{figure}
\begin{center}
\includegraphics[width=10cm]{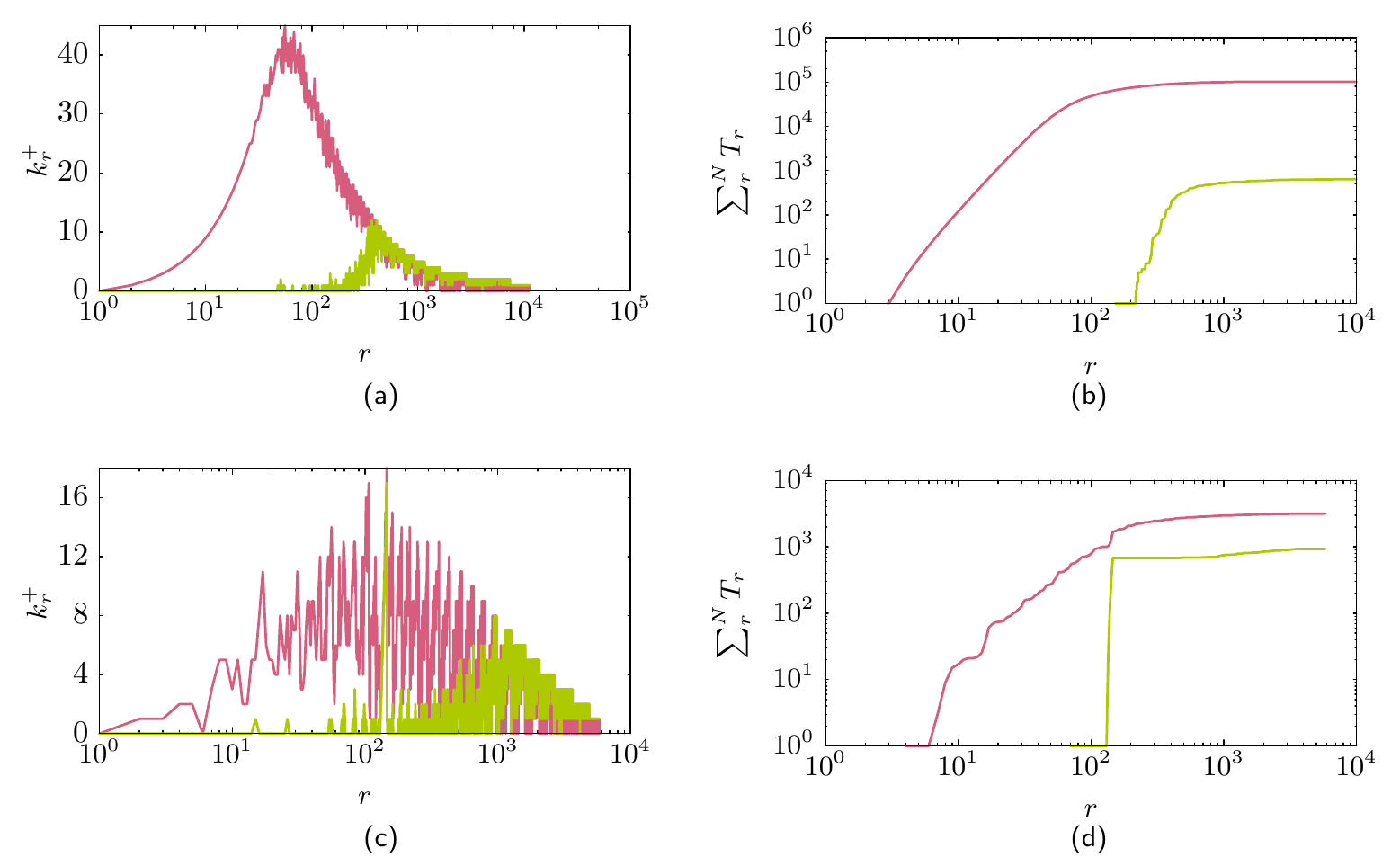}
\end{center}
\caption{\label{fig:hubsTria} 
(a) Difference in the hub connectivity for two networks that have the same degree distribution as the AS-Internet. One network has maximal connectivity within the nodes (pink line) ($\rho=-0.181$) and the other minimal connectivity within the hub nodes (green line) ($\rho=-0.20$). (b) This change of connectivity is reflected in the number of triangles in each network, where the network with maximal hub connectivity (pink line) has almost two orders of magnitude more triangles that the other network (green line). (c)-(d) Similar as (a)--(b) but for the Hep--Th network where assortativity coefficients $\rho = 0.69$ and $\rho = -0.42$.
}
\end{figure}

The connectivity within the hubs has a significant impact in the number of triangles in a network. Fig.~\ref{fig:hubsTria}(a) shows $\linksRC_i$ for two networks. Both networks have the same degree distribution as the AS--Internet but one network has the maximal possible connectivity within the  hubs (pink) and the other network has the minimal possible connectivity within the hubs (green). The networks were created using restricted randomisation so there are no multilinks. Fig.~\ref{fig:hubsTria}(b) shows the cumulative number of triangles $\sum_{r=1}^N T_r$ for these networks as the rank increases. Notice that the difference is almost two orders of magnitude between the network with maximal hub connectivity against the one with minimal hub connectivity. However both networks have similar assortativity coefficient due to the structural correlations as for both of these networks $\degreeMax>\sqrt{N\langle k \rangle}$.

Figures~\ref{fig:hubsTria}(c)--(d) show the case of two networks which have the same degree distribution as the Hep--Th. Again the difference between the networks is the connectivity within the hubs, and again as in the previous case, decreasing the connectivity of the hubs decreases the number of triangles in the network. In this case there are no structural constraints and the two networks have very different assortativity coefficient. The network with maximal connectivity of the hubs is assortative ($\rho=0.69$) compare with the other network which is dis-assortative ($\rho=-0.42$). This is an example where the change on the connectivity of the hubs has a drastic effect on the assortativity coefficient and the number of triangles in the network.

\section{\label{sec:cores}Cores}

The degree sequence gives a centrality measure to distinguish the nodes. It is common to assume that nodes of higher degree are more important and form the core of the network. There are several possibilities to define a core via the sequences $\{\linksRC_i\}$ and  $\{K^+_i\}$. 

\subsection*{The rich--core}
One of the simplest ways to define a core is via the rich--core~\cite{ma2015rich}.  The core is a set of well connected nodes, that is the top $r_c$ ranked nodes, the periphery the rest. The boundary of the rich--core is the rank $r_c$ where $\linksRC_{r_c}$ is maximal.  
The core are all the nodes with rank less than $r_c$. 

\begin{figure}
\begin{center}
\includegraphics[width=8cm]{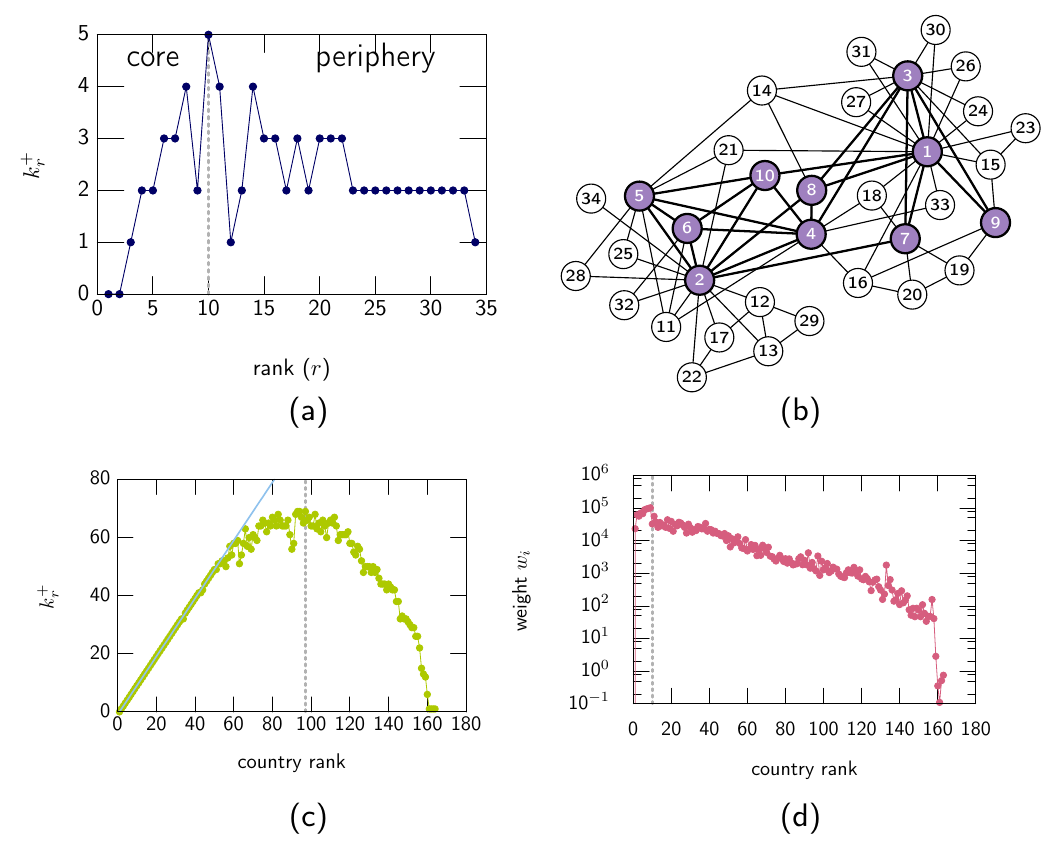}
\caption{\label{fig:defRichCore}
(a) Rich--core for the Karate network defined by the maximum of  $\linksRC_r$  at $r=10$. The members of the core are the top 10 ranked nodes as shown in (b). (c) Example of $\linksRC_r$ of the trade relationship between WTO countries in 1980. The core of this network is the top 97 ranked nodes (dash vertical line). The diagonal line (light blue) shows the value where the top ranked nodes form a clique. (d) The weighted $\weightPlus$ for the WTO countries where $\weightPlus$ is related to the trade in dollars. The core is the top ten ranked nodes (dash vertical line). Notice that the vertical axis is in logarithmic scale. 
}
\end{center}
\end{figure}

Figure~\ref{fig:defRichCore}(a) show an example of this core for the Karate club. The maximum value happens at $r_c=10$ so the top 10 nodes form the core, the core is shown in Figures~\ref{fig:defRichCore}(b).  The partition of a network via the properties of $\linksRC_i$ is attractive due to its simplicity and this partition has been used to define the core of the \emph{C. elegans} and its time evolution~\cite{ma2015rich}, in the characterisation of food webs~\cite{lu2016drought} and recently in the concept of rich--core has been extended to multiplexes when studying the brain connectivity~\cite{battiston2018multiplexA}.

The rich--core can also be evaluated in weighted networks. Figures~\ref{fig:defRichCore}(c)-(d)  show the rich--core of two networks describing the trade between nations in 1980 as reported by the World Trade Organisation (WTO). In figure~\ref{fig:defRichCore}(c) shows $\linksRC_r$ which in this case corresponds to the number of  trade relationships that country $r$ has with countries that have at least the same amount of trade relationships as country $r$. The figure also shows the line $r-1$, this line is the maximum amount of trade relationships that country $r$ can have with countries of higher rank. It is clear that the top 50 countries form an almost fully connected clique. The core is formed by 97 countries which is almost 60\% of total number of countries reported in the WTO dataset for 1980. 
Figure~\ref{fig:defRichCore}(d) shows $w_r^+$, the trade relationship weighted by the amount of dollars (exporting goods) that country $r$ has with countries that are larger exporters in value than itself. In this case the core is formed by only 10 countries, that is around 6\% of the countries.  The cores defined by  $\linksRC_r$ and $w_r^+$ show two different views of the trade between  nations. There is a large number of trade agreements between nations and a large number of nations are part of the core of these agreements, however by value less than 10 nations form the core. 

\begin{figure}
\begin{center}
\subfigure[]{
\includegraphics[height=4cm]{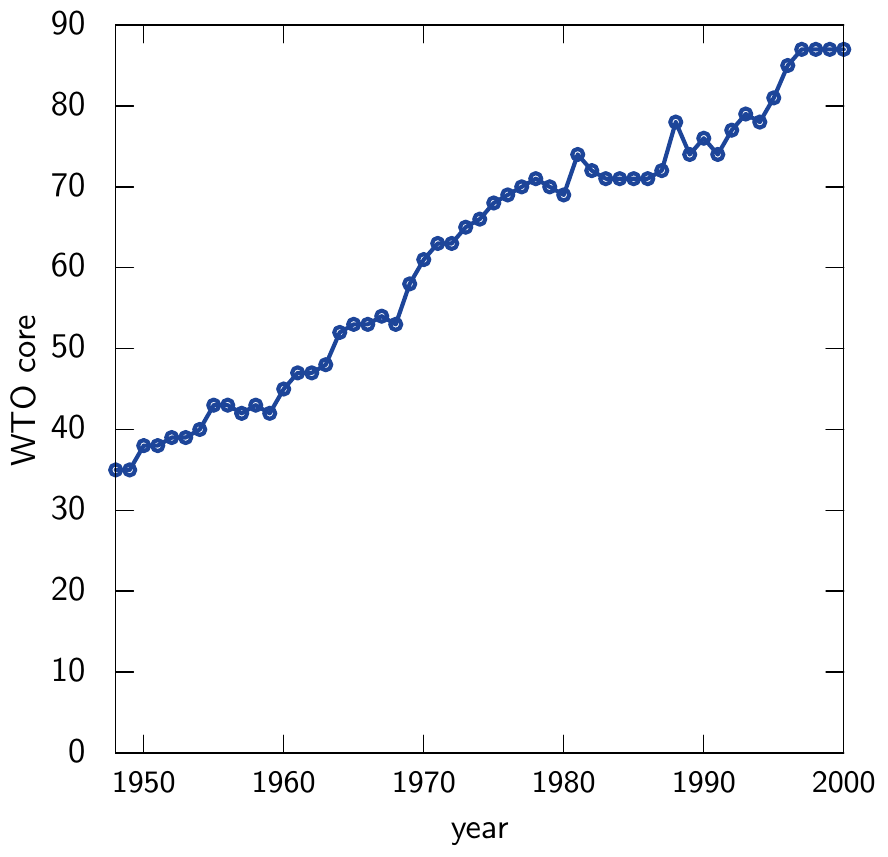}
}
\subfigure[]{
\includegraphics[height=4cm]{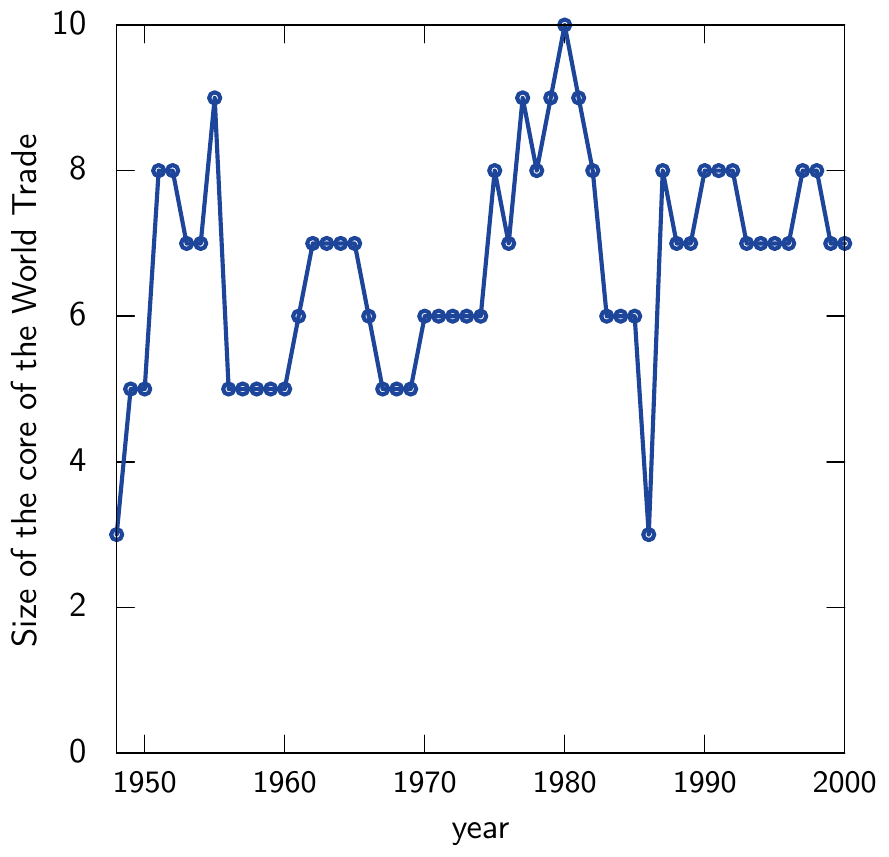}
}
\caption{\label{fig:usesRichCore}
Evolution of the rich--core by  (a)  trade relationships between countries and (b) by value of exports between countries. }
\end{center}
\end{figure}

The evolution of the hubs can be described via the rich--core. Figure~\ref{fig:usesRichCore}(a)-(b) shows the evolution of the cores from 1950--2000 for the trade relationships and weighted relationships between countries.  The amount of trading nations has increased from 1950--2000 (Fig.~\ref{fig:usesRichCore}(a)), perhaps due to globalisation and the core of trading nations has become larger with time. 
However by wealth (Fig~\ref{fig:usesRichCore}(b)), the size of the core has not changed much, less than ten countries dominate the market by value. 

\subsection*{The spectral--core}
Another common measure of centrality is the eigenvector--centrality. In this case a node is important if it is connected to other important nodes.  It is know that in many networks the degree centrality of a node is correlated to its eigenvector centrality so it is natural to ask what is the contribution of the nodes with high degree centrality to their eigenvector centrality.

For an undirected and unweighted network whose connectivity is described by the adjacency matrix ${\bf A}$, the spectrum of the network is the set of eigenvalues $\firstEigenvalue\ge \Lambda_2\ldots \ge \Lambda_N$ of ${\bf A}$. The highest eigenvalue $\firstEigenvalue$ plays an important role when describing information diffusion or epidemic transmission on a network~\cite{gomez2010discrete,van2011n,wang2003epidemic,youssef2011individual}. 

It is known that the spectral radius $\firstEigenvalue$ increases with the assortativity coefficient and it is related to the number of triangles in the networks~\cite{van2010influence,d2012robustness,estrada2011combinatorial}, so it is expected that changes in the hubs connectivity would also change the spectral radius. For the eigenvalue $\firstEigenvalue$, its corresponding eigenvector $\firstEigenvector$ is the eigenvector centrality where the entry $(\firstEigenvector)_i$ is the ``importance'' of node $i$.

The sequence $\{\linksRC_i\}$ can be used to define a lower bound for $\Lambda_1$~\cite{mondragon2016network}
\begin{equation}
\Lambda_1\ge 2\langle \linksRC \rangle_{{r}}=\frac{2}{r}\sum_{i=1}^r \linksRC_i
\end{equation}
where the $\langle\linksRC\rangle_r$ is the average number of links shared by the top $r$ nodes. These bound can be used to define a core. The spectral--core  boundary is the rank $r_c$ where $\langle \linksRC \rangle_{r_c}$ is maximal. This correspond to the best bound of $\firstEigenvalue$ based on $\linksRC_i$. Similarly as the rich--core, any node with rank less than $r_c$ belong to the core.

This bound can also be used to create an approximation of the eigenvector centrality $\firstEigenvector_1$.  If the core is defined by the rank $r_c$ then an approximation to $\firstEigenvector_1$ is the vector $\overline{y}$ with entries $y_i=K_i^+(r_c)$ where $K_i^+(r_c)$ is the number of links that node $i$ shares with the top $r_c$ nodes.

The bound of $\firstEigenvalue$ can be improved if we consider the average of  links which have at least one of its end nodes in the core
\begin{equation}
\label{eq:spectralCoreCond}
h(r) =\frac{1}{r}\sum_{i=1}^N K^{+}_i(r)
\end{equation}
which is the average number of links that connect to the top $r$ nodes.
The core boundary is defined by the value of $r$ when $h(r)$ is maximal. The bound in this case is $\sqrt{h(r_c)}\le \firstEigenvalue$.
It is also possible to construct an approximation to the eigenvector centrality from this bound. If $\walk_2(r_c, i)$ are the number of walks of length two that start in one of the $r_c$ top nodes and end up in any other node $i$ then the approximation $\overline{y}$ to the eigencentrality has entries $y_i=\walk_{2}(r_c,i)$. Interestingly in this approximation centrality we not only consider the nodes that form the core but also nodes that connect directly to the core. 

As an example we show the spectral core for the EU--Air transportation networks. The dataset consist of the 37 airlines that connect 450 different European airports. First we consider a simple version of this network. The nodes represent different EU airports and a links represent if there is a connection between two airports via a flight. We do not consider the number of flights between two airports or differentiate the airlines.

\begin{figure}
\begin{center}
\includegraphics[width=11cm]{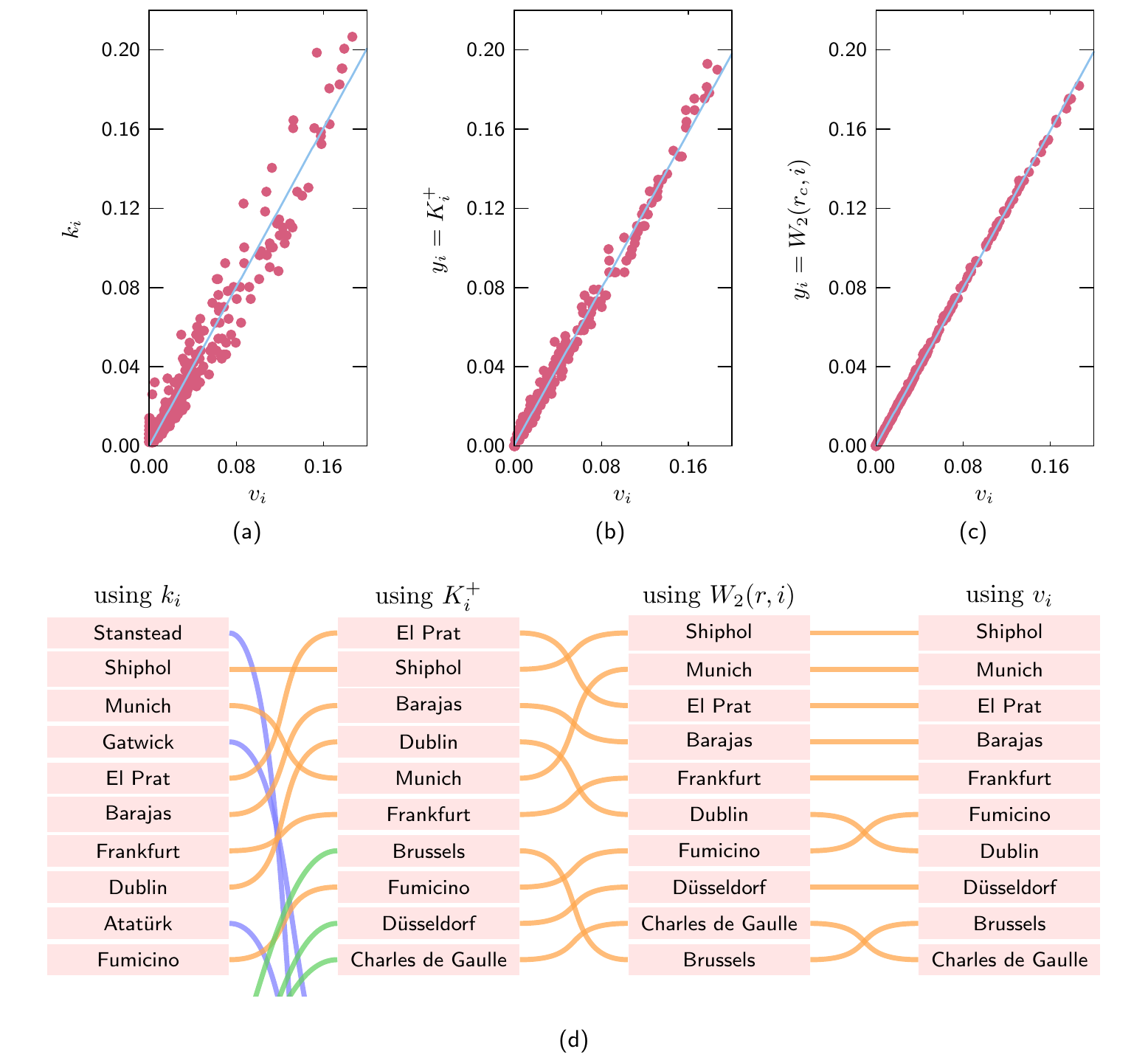}
\caption{\label{fig:defSpectralCore}
Correlation between the eigencentrality and its approximations for the EU Air transportation network. (a) Approximation using the degree, (b) $y_i=K_i^+(r_c)$ and (c)  $y_i=\walk_{2}(r_c,i)$. (d) Change on the ranking of the top nodes when ranked by decreasing order of their degree $\degree_i$, $y_i=K_i^+(r_c)$, $y_i=\walk_{2}(r_c,i)$ and the eigencentrality $v_i$. 
}
\end{center}
\end{figure}

Figure~\ref{fig:defSpectralCore} shows the approximation of the eigencentrality via the degree of the nodes (Fig.~\ref{fig:defSpectralCore}(a)),  the number of links connecting to the core $y_i=K_i^+(r_c)$ (Fig.~\ref{fig:defSpectralCore}(b)) and the number of walks of length one that finish in the core $y_i=\walk_{2}(r_c,i)$ (Fig.~\ref{fig:defSpectralCore}(c)). 
The size of the spectral--core is 82 nodes. The spectral core based on the walks of length two gives a good approximation to the eigencentrality. Figure~\ref{fig:defSpectralCore}(d) shows the airports ranked in decreasing order of the different centralities.

We finish this section with two observations. The size of the core is not related to the assortativity of the network, it is possible to have networks with small core which are disassortative, e.g. the AS--Internet. The other observation is that in networks that are highly disassortative the approximation to the largest eigenvalue (Eq.~(\ref{eq:spectralCoreCond})) can be good, however this not translates to a good approximation to the eigencentrality (see Appendix for an example).
	
\subsection*{Biased-random walk core and centrality}
Random walks on networks are used to understand structural properties of networks like community detection~\cite{rosvall2008maps}, centrality of nodes~\cite{noh2004random}, discovery of the network structure~\cite{yoon2007statistical} and the partition of a network into  core--periphery~\cite{della2013profiling}. 
The maximal entropy random walk~\cite{burda2009localization} (MERW) has the property that the random walker would visit walks of same length with equal probability. This kind or random walks have been used to study different properties of complex networks~\cite{li2011link,ochab2013maximal} and in some applications~\cite{yu2014maximal,leibnitz2013maximum}.

In the Maximal Entropy Random Walk (MERW) the transition probability from node $i$ to node $j$ is~\cite{burda2009localization} 
\begin{equation}
\label{eq:MERW}
\jumpProb=\frac{({\bf A})_{ij}v_j}{\sum_j ({\bf A})_{ij}v_j}=\frac{({\bf A})_{ij}v_j}{\firstEigenvalue v_i},
\end{equation}
where $({\bf A})_{ij}$ is the $i$, $j$ entry of the adjacency matrix and $v_i$ is the $i$--th entry of the eigenvector centrality $\firstEigenvector$. 
The stationary probability $p^*_i$, which is the probability of finding the walker in node $i$ as time tends to infinity,  for the MERW is $p^*_i=v_i^2$.%

If the largest eigenvalue-eigenvector pair is not known an approximation to the the MERW  can be obtained from an approximation to the eigenvector $\firstEigenvector$. As mentioned in the spectral--core section, $y_i(r)=K^+_i(r)$,  gives an approximation to the eigenvector $\firstEigenvector$.

A biased random walk based on this approximation is~\cite{mondragon2018}
\begin{equation}
P_{i\rightarrow j} =  \frac{({\bf A})_{ij}(K^+_j(r)+1)}{\sum_j^N ({\bf A})_{ij}(K^+_j(r)+1)}.
\label{eq:biasedCore}
\end{equation}
The term 1 in the numerator and denominator is added as it it is possible that $K^+_j(r)=0$ if node $j$ has no links with node of rank greater than $r$ and then the random-walk will be ill-defined. This core--biased random walk describes the dynamics of  a random--walker which prefers to jump to the hubs that have many connection with other hubs. 

The MERW has the property that in networks where hubs are present,  the stationary probability $p^*_i=v_i^2$ of the hubs is relatively large,  there is an argument~\cite{martin2014localization,lin2018non} saying that the property of concentrating the eigencentrality in the hubs is an undesirable property as diminishes the effectiveness of the centrality as a tool for quantifying the importance of nodes.

The core--biased random walk is based on the hubs,  the relevant hubs are the ones that are well connected with other hubs, this is reflected in the stationary probability $p^*_i$ which we can consider as a centrality measure based only in the interconnectivity of the hubs (see Appendix for the evaluation of $p^*_i$).
Fig.~\ref{fig:biasedCentrality}(a)-(c) shows the  plots of the network-scientist network where fig.~\ref{fig:biasedCentrality}(a) shows the layout of the network, fig.~\ref{fig:biasedCentrality}(b) shows the same layout but with the radius of the nodes proportional to the eigenvector centrality i.e. $p^*_i=v_i^2$. As noticed before the eigencentrality is concentrate in the hub nodes and diminishes the importance of nodes of low degree. Fig.~\ref{fig:biasedCentrality}(c) shows the  network with the radius proportional to the stationary probability obtained by the core-biased random walk. In the figure the orange nodes are the core of the network and now nodes of low degree are part of the core as they are locally important.
\begin{figure}
\subfigure[]{\includegraphics[width=4.5cm]{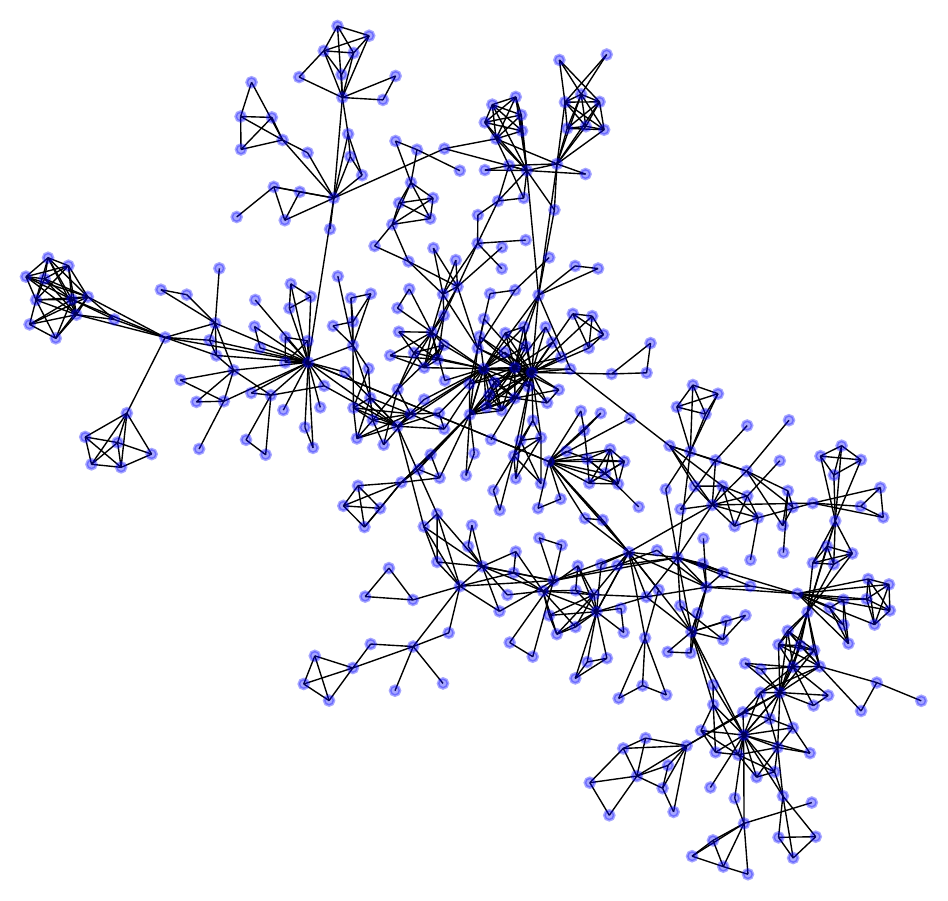}}
\subfigure[]{\includegraphics[width=4.5cm]{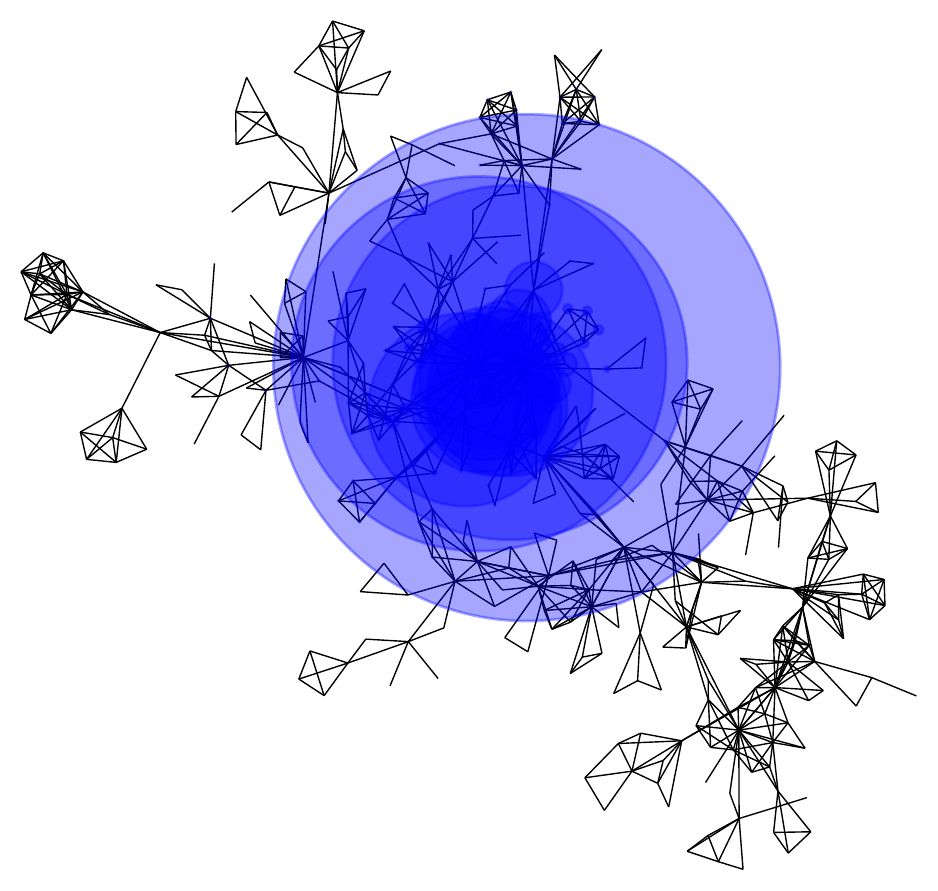}}
\subfigure[]{\includegraphics[width=4.5cm]{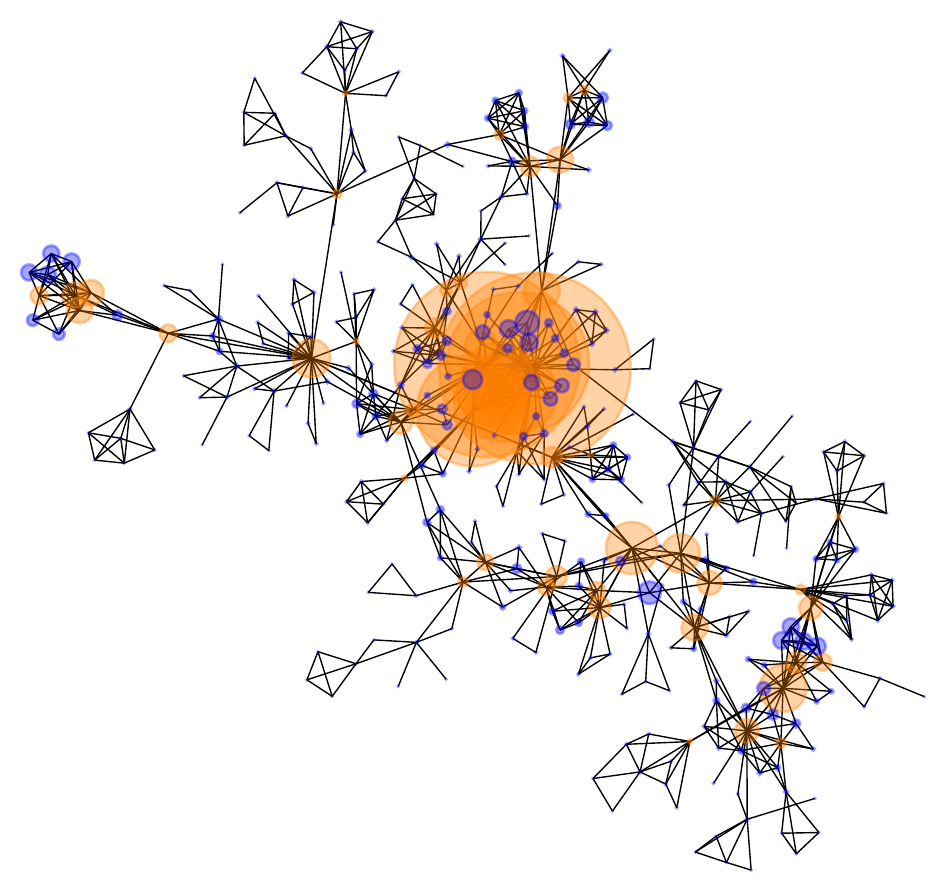}}
\caption{\label{fig:biasedCentrality}
(a) Graph of the Network Scientist network.
(b) The same graph but the radius of the nodes is proportional to $p^*_i=v_i^2$ where $v_i$ are the entries of the eigencentrality.
(c) In this case the radius of the nodes is proportional to the stationary probability $p^*_i$ of the core--biased random walk. The orange nodes are the core of the network. The core is the top ranked  54 nodes. 
}
\end{figure}

\subsection*{The rich--club}

The measure which describes how tightly the rich--nodes are interconnected is  the rich-club coefficient~\cite{Zhou04}. This coefficient is the ratio between the number of connections that the rich--nodes have against the maximum number of connections that they could have, that is the density of connections within the hubs which can be expressed in terms of $\linksRC_i$ as
\begin{equation}
\label{eq:RichClubCoefRank}
\richClubRank(r) = \frac{2\sum_{i=1}^{r}\linksRC_i}{r(r-1)},
\end{equation}
where the sum is the total number of links shared by the top $r$ nodes and the factor $(r(r-1))/2$ is the total number of links that can exist between these nodes. It is also common to define  the rich--club coefficient in terms of the degree~\cite{Colizza06}, in this case,
\begin{equation}
\richClubDegree(k)=\frac{2E_{>k}}{N_{>k}(N_{>k}-1)}
\end{equation}
where $E_{>k}$ is the number of links shared between the nodes of degree greater than $k$ and $N_{>k}$ is the number of nodes that have degree greater than $k$. Notice that the rank based and degree based rich-clubs are related by $\richClubRank(r_{>k})=\richClubDegree(k)$,
where $r_{>k}$ is the node with lowest rank and degree greater than $k$ then $E_{>k}=\sum_{i=1}^{{r_{>k}}}\linksRC_i$ and $r_{>k}=N_{>k}$.
The rich--club coefficient and its generalisations have been proved to be a
useful measure for studying complex networks~\cite{Colizza06,McAuley07,Opsahl2008,serrano2008rich,Zlatic09,alstott2014unifyingA}.
In recent years it has been used to describe the connectivity of the brain, the connectome~\cite{Heuvel11,10.3389/fnhum.2014.00647,ball2014rich,grayson2014structural,nigam2016rich,collin2013impaired,senden2014rich}. 

Originally the rich--club was defined as the set of nodes that are tightly connected~\cite{Zhou04}, that is the set of nodes where the rich--club coefficient is or tends to 1. A clique would have a rich--club coefficient of 1. Colizza~et~al.~\cite{Colizza06} introduce an alternative definition.
They defined the rich--club as the set of high degree nodes that have a density of connections higher than expected. The expected number of links between the top ranked nodes is evaluated from a random network which is considered as a null model. If $\richClubDegree_{\rm rand}(k)$ is the rich-club coefficient of a random network which has the same degree sequence as the original network then the comparison between the network and the null model is done via the normalised rich-club coefficient $\richClubDegree_{\rm norm}(k) =\richClubDegree(k)/\richClubDegree_{\rm rand}(k)$.
Colizza et~al.~\cite{Colizza06} stated that an indicator of a rich-club with respect to the null-model is when  the normalised rich-club coefficient is greater than 1. 

\begin{equation}
\richClubRank_{\rm norm}(r) = \frac{\richClubRank_{\rm norm}}{\richClubRank_{\rm rand}}=\frac{\sum_{i=1}^{r}\linksRC_i}{\sum_{i=1}^{r}\kappa^+_i}%\overline{\linksRC_i}}
\end{equation}
where $\kappa_i^+$ denotes  the connectivity of node $i$ with nodes or larger rank obtained from one of the ensembles or by restricted randomisation.  If the maximal degree present in the network is less than the structural cut--off degree  then the uncorrelated rich--club can be approximated via 
\begin{equation}
\richClubRank_{\rm rand}(r) =
\frac{2\alpha}{r(r-1)N\langle k \rangle}\sum_{i=1}^{r}\degree_i \sum_{n=1}^{i-1}\degree_n
\end{equation}
and the normalised  rich-club is
\begin{equation}
\richClubRank_{\rm norm}(r) = \frac{N\langle k \rangle \sum_{i=1}^r \linksRC_i}{\sum_{i=1}^r\degree_i\sum_{n=1}^{i-1}\degree_n}.
\end{equation}

\subsection*{Changes related to the core connectivity}
The change of connectivity between the hubs could happen due  to the disappearance and/or the reshuffling of some of its links, here we consider that some links between hubs are removed. As the cores are defined via the connectivity of the hubs, any change in this connectivity would result in a change of which nodes belong to the core. 

The changes in the assortativity coefficient would be constraint if the maximal degree is larger than the cut--off degree, if this is the case,   changes in the hubs connectivity would have little effect on the degree--degree correlations.  The changes of the hub connectivity  have a large impact in the number of triangles and longer loops in the network. Recently, it was suggested~\cite{10.3389/fnhum.2014.00647} that a good measure to quantify the changes of the network connectivity is to use the \emph{communicability}~\cite{estrada2008communicability} of a network. 

The {communicability} between nodes $p$ and $q$ is defined via the weighted sum of all walks between $p$ and $q$ as
\begin{equation}
G_{pq}=\sum_{k=0}^\infty \frac{\left( {\bf A}^k \right)_{pq} }{k!}=\sum_{j=1}^N \overline{v}_j^{(p)}\overline{v}_j^{(q)} e^{\Lambda_j}
\end{equation}
where $\Lambda_p$ is the $p$-th eigenvalue and $\overline{v}_j^{(p)}$ is the $j$ entry of the eigenvector $p$-th eigenvector. The communcability has the property that it is affected by structural changes of the walks.

\begin{figure}
\begin{center}
\subfigure[]{\includegraphics[width=4.5cm]{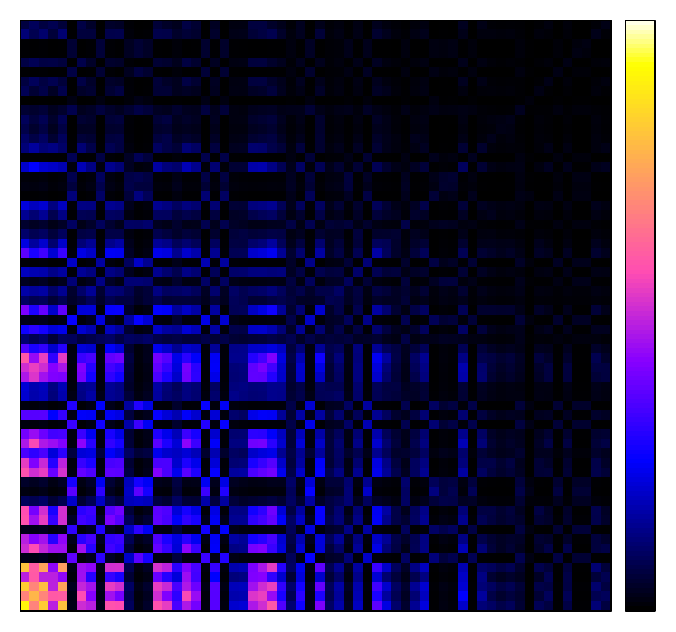}}
\subfigure[]{\includegraphics[width=4.5cm]{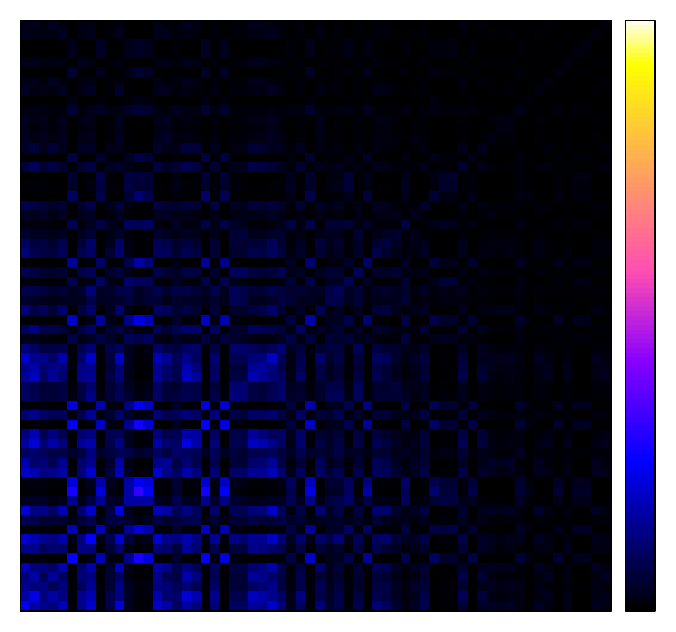}}\\
\subfigure[]{\includegraphics[width=4.5cm]{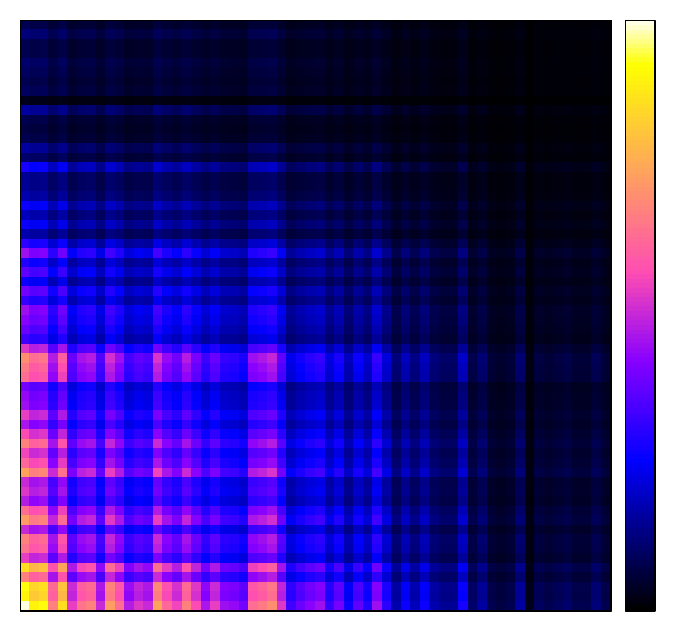}}
\subfigure[]{\includegraphics[width=4.5cm]{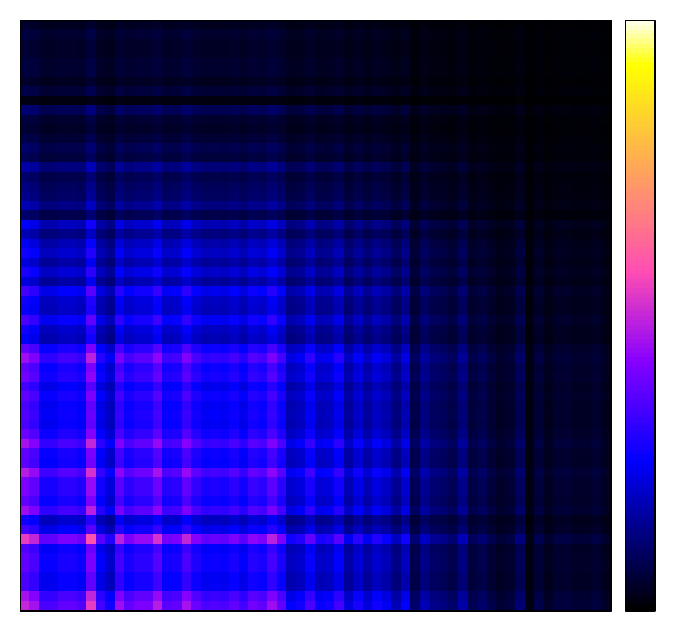}}
\end{center}
\caption{Communicability for (a) the original Dolphins network and (b) when the links between the top 10 nodes are removed. (c)--(d) Approximation to the communicability shown in (a) and (b)  using the  approximation to the top eigenvalue-eigenvector pair from the spectral--core.
\label{fig:commuDolphins}
}
\end{figure}

Fig.~\ref{fig:commuDolphins}(a)--(b) shows the communicability of the Dolphins network for the original dataset (Fig.~\ref{fig:commuDolphins}(a)) and for the network where the links between the top 10 ranked  nodes are removed (Fig.~\ref{fig:commuDolphins}(b)). Clearly in this case, the communicability strongly depends on the connectivity within these 10 hubs.  This dependence is also capture in the approximation of the communicability using only the properties of the hubs. 
Figs.~\ref{fig:commuDolphins}(c)--(d) show the approximation of communicability using  the approximation of the eigenvalue--eigenvector pair from the spectral-core based on walks of length two (Eq.~(\ref{eq:spectralCoreCond})).
 
\subsection*{Evolution of a core}
If the connectivity within the well connected nodes has such a large influence in the degree--degree correlation, how are these cores created?   Recently Fire and Guestrin~\cite{fire2017rise} studied how the rich nodes appear and disappear in networks. They studied the evolution of 38,000 networks and  noted that the creation of nodes of high degree is correlated to the speed of growth of the network. In slow growing networks the hubs appear shortly after the network becomes active and if a node becomes a hub it tends to stay a hub. In fast growing methods the hubs can appear at any time in the network evolution and their position as a top hub can change as new hubs with higher degree can appear as the network evolves. 

The Barabasi--Albert (BA) model based on preferential attachment is an example of what Fire and Guestrin classify as a  slow growing network. The BA  model creates links between a new node and old nodes and can produce nodes with high degree. The degree of a hub is correlated with its age, nodes that becomes hubs at early stages of the network evolution, tend to remain a hub. The BA model will not produce hubs that are well interconnected, as  the network evolves the \emph{rich-gets-richer} mechanism increases the connectivity of the hubs but do not increase the connectivity within the hubs. It is known that network models based on the addition of new links between old nodes can have drastic changes in the overall structure of the network~\cite{Simon55,Bornholdt01,dorogovtsev2010lectures}.  In the case that the addition of new links is biased towards connecting the hubs, i.e.   the rich-club phenomenon~\cite{Zhou04}, then the network would have a well connected core. 

\begin{figure}
\begin{center}
\includegraphics[width=5cm]{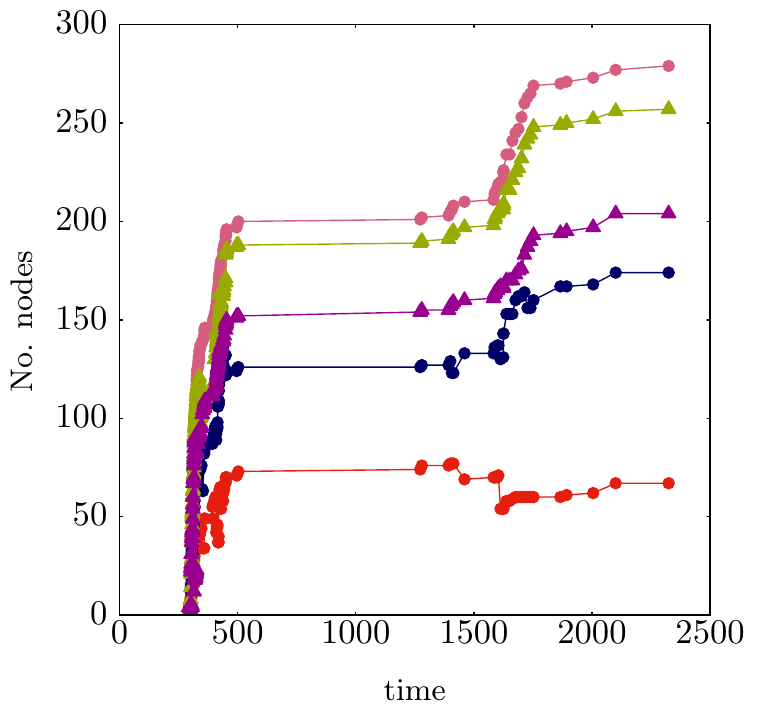}
\end{center}
\caption{\label{fig:celegnasGrowth} Growth of the number of neurones for the \emph{C. elegans}. The top curve is the total number of neurones. From bottom to top, the rich--core, the biased--random walk, the spectral core based on walks of length two, the spectral core based on the hubs connection density and the number of neurones (pink).  }
 \end{figure}

An example of a non-trivial evolution of a network and its cores is in Fig.~\ref{fig:celegnasGrowth} which shows the number of neurones of the \emph{C. elegans} from birth to maturity. The cores based on spectral properties of the network tend to grow with the network. The rich--core diminishes in size in the second spurt of growth around the 1500 time mark. The top ranked neurones are born between the 250 to 450 time mark, but there are exceptions, for example the neurone PVR is born in the 2100 time mark and ends ranked on the top 33 neurones. Around the second spurt of growth the well connected neurones increase their interconnectivity, reducing the size of the rich-core and increasing the spectral related cores. The growth of these cores correspond to an increase on the spectral radio $\Lambda_1$ and an increase on the number of triangles in the network.

There are some network growth models specifically designed to reproduce the connectivity within the well connected nodes~\cite{zhou2008pfp,csigi2017geometric}. However more research needs to be done to understand other mechanisms related to the formation of cores, in particular for fast growing networks.

\section{Discussion}
The description of a network using the degree and the connectivity with the better connected, i.e. the sequences $\{\degree_i\}$ and $\{\linksRC_i\}$,  produces networks with similar correlations as the original network with the advantage that the statistical description of properties related to high degree nodes are well defined even for power law networks.
The connectivity within the well connected can have a large effect on the assortativity and clustering coefficient. In the case that the network maximal degree is larger than the structural cut--off degree, changing the connectivity within the hubs has a little effect on the correlations. However it has a large effect on the number of triangles in the network. For networks that are well modelled by the degree distribution and degree--degree correlations, ensembles based on the sequences $\{\degree_i\}$ and $\{ \linksRC_i\}$ give a good approximation to these networks.

The core of a network can be defined using the connectivity within the well connected $\{\linksRC_i\}$ or with the well connected $\{K^+_i\}$. The spectral and random--bias cores are based on an approximation of the spectral radius from the connectivity related to the hubs. These cores are related to the eigenvector centrality and can be used to define centrality measures based on the hubs relative importance.  The ensembles and the cores are related as the degree--degree correlation, the clustering and the spectral radius are all related to the connectivity of the hubs, confirming the importance that the hubs play in the overall structure of the network. 

%\bibliographystyle{abbrv}
%\bibliography{BeyondRich.bib}

%\putbib
%\end{bibunit}
\newpage

%\begin{bibunit}
\section*{Appendix. Ensembles based on the rich--nodes connectivity}
\subsection*{Maximal entropy approach}
Consider a network described by the sequences $\{\degree_1 ,\ldots,\degree_N\}$ and $\{\linksRC_1,\ldots, \linksRC_N\}$ where $N$ is the number of nodes, $L$ is the number of links and self--loops are not allowed. These sequences satisfy $L = \sum_{r=1}^N \degree_r=2\sum_{r=1}^N\linksRC_r$. Here we are assuming that the nodes have been ranked in decreasing order of their degree.  From these sequences it is possible to construct an ensemble (or a null--model) using the Maximal Entropy approach~\cite{Mondragon2014}.

The Shannon entropy of the network is $\entropy = -\sum_{i=1}^N \sum_{j=1; j\ne i}^N \eProb_{ij} \log( \eProb_{ij} )$. The maximal entropy is the set of probabilities where the entropy $\entropy$ is maximal under certain constraints. Here the constraints are the normalisation, $\sum_i\sum_j\eProb_{ij}=1$, the conservation of $\linksRC_r$
\begin{equation}
\sum_{i=1}^{r-1}\eProb_{ir}=\frac{\linksRC_r}{L}, \quad r=1,\ldots ,N-2
\end{equation} 
and the conservation of $\degree_r$
\begin{equation}
\sum_{i=1}^N \eProb_{ir}=\frac{\degree_r}{L}=\frac{\linksRC_r}{L}+\sum_{i=r+1}^Np_{ir},\quad r=1,\ldots , N-1.
\end{equation}

The common procedure to obtain the maximal entropy solution is first to label the links via the nodes labels $i$, $j$ via the map $\ell=g(ij)$ and then transform $\eProb_{ij}=\eProb_\ell=e^{q_\ell}$. The constraints are expressed as $\sum_{\ell}^{N(N-1)/2}f_m(\ell)e^{-q_\ell}=c_m$ where $c_m$ are $M$ constraints that are related to $q_\ell$ via the map $f_m(\ell)$. The solution of the maximal entropy under the constraints is obtained using the Lagrangian multipliers $\lambda_0, \ldots,\lambda_M$ and the maximisation of $\mathcal{F}(q_1,\ldots,q_{N(N-1)/2})=\sum_\ell^{n(N-1)/2}(q_\ell+\lambda_0)e^{-q_\ell} + \sum_m^M\lambda_m\sum_\ell^{n(n-1)/2}f_m(\ell)e^{-q_\ell}$. The maximisation of $\partial\mathcal{F}(q_1\ldots)/\partial q_\ell=0$ for $\ell=1,\ldots,N(N-1)/2$ gives the solution 
\begin{equation}
\label{eq:maxentOne}
q_\ell=1-\lambda_0\sum_{m=1}^M\lambda_mf_m(\ell).
\end{equation} 
This last equation combined with the constraint equations are solved to obtain the MaxEnt solution. 
 Usually the solution of the MaxEnt is evaluated using the Partition function formalism which  gives a smaller set of non-linear equations to solve. However, for the case that the constraints are the sequences $\{\degree_i\}$ and $\{\linksRC_i\}$ the solution can be obtained directly from Eq.~(\ref{eq:maxentOne}) and the constraint conditions. 
The Maximal Entropy solution is given by the probabilities~\cite{Mondragon2014}
\begin{equation}
\label{eq:mainResult}
\eProb_{ij} = 
\frac{\stepFunct(i)\left(\degree_i-\linksRC_i \right) }{\sum_{n=1}^{j-1} \stepFunct(n)\left(\degree_n-\linksRC_n \right)}\frac{ \linksRC_j}{L}\quad i<j\\
\end{equation}
where 
\begin{equation}
\label{eq:stepFunct}
\stepFunct(n) = \frac{\stepFunct(n-1)\sum_{i=1}^{n-1}\stepFunct(i)(\degree_i-\linksRC_i)}{\sum_{i=1}^{n-1}\stepFunct(i)(\degree_i-\linksRC_i)-\linksRC_n\stepFunct(n-1)}.
\end{equation}
The values of $\stepFunct(n)$  are defined recursively with the initial condition $\stepFunct(1)=1$. The average number of links between nodes $i$ and $j$ is $e_{ij}=Lp_{ij}$ with variance ${\rm var}(e_{ij})=Lp_{ij}(1-p_{ij})$.
By construction the ensemble satisfies the `soft' constraints $\langle \degree_r\rangle = \sum_{j=1}^N L \eProb_{rj}=\degree_r $ and $\langle \linksRC_r\rangle = \sum_{j=1}^{r-1} L\eProb_{rj}=\linksRC_r $, where the angled brackets denote expected value. The variance of the degree is $\sigma^2(\degree_r)=L\sum_{j\ne r}^Np_{rj}(1-p_{rj})$.  

In the following sections the ensembles ME1, ME2 and ME3 are the ones defined in the main manuscript.

\subsection*{Correlations}
 To characterise the networks produced by the ensembles we used two measures, the average neighbours degree and the assortativity coefficient. 
\subsubsection*{Average neighbours degree}
The average nearest neighbours degree given by $\langle \averageDegDeg(k) \rangle =\sum_{k'} k' P(k'|k)$~\cite{pastor2001dynamical}, where $P(k'|k)$ is the conditional probability that given a node with degree $k$ its neighbour has degree $k'$. 

In our case $\eProb_{ij}$ is the probabilities obtained from on of the ensembles then
\begin{equation}
\langle \averageDegDeg(k)\rangle = \frac{1}{N_k}\sum_{i=1}^{N}\left(\frac{1}{k}
\sum_{j=1}^{N} \eProb_{ij}L k_j\right)\delta_{k_i,k},
\label{eq:knn}
\end{equation}
where $\delta_{k_i,k}=1$ if $k_i=k$ and zero otherwise.

\subsubsection*{Assortativity coefficient}
The assortativity is evaluated using~\cite{Newman02}
\begin{equation}
\rho =\frac{\left\langle \degree \degree' \right\rangle_\ell- \left\langle k \right\rangle^2_\ell}{\left\langle k^2\right\rangle_\ell-\left\langle k  \right\rangle_\ell^2}
\end{equation}
with
\begin{equation}
\left\langle \degree \right\rangle_\ell=\sum_i\sum_{j\ne i} \degree_i p_{ij}=\frac{\langle k^2 \rangle_n }{ \langle k \rangle_n }
\end{equation}
where $\langle \ldots\rangle_\ell$ is the average over all links and $\langle \ldots \rangle_n$ is the average over all nodes.
The average degree of the end nodes of a link is $\left\langle k k' \right\rangle_\ell = \sum_i\sum_{j\ne i} \degree_i \degree_j p_{ij}$. 
Then 
\begin{equation}
\rho = 2\frac{T_1-T_2T_2}{T_3-T_2T_2}
\end{equation}
where $T_1=\langle k k'\rangle=\sum_{i=1}^N\sum_{j=1}^N p_{ij}k_i k_j$, $T_2=\langle k\rangle = \sum_{i=1}^N\sum_{j=1}^N p_{ij}(\degree_i+\degree_j)/2$ and $T_3=\langle k^2 \rangle=\sum_{i=1}^N\sum_{j=1}^N p_{ij}(\degree_j^2+\degree_i^2)/2$. 

\subsubsection*{Some observations}
Different networks can have the same assortativity coefficient $\rho$ or average neighbours degree $\langle \averageDegDeg(k)\rangle $ but different $\linksRC_i$, so there is no simple relationship between the
density of connections between the rich nodes and these two measures of correlation. 

\begin{figure}
\begin{center}
\subfigure[]{
\includegraphics[width=4.5cm]{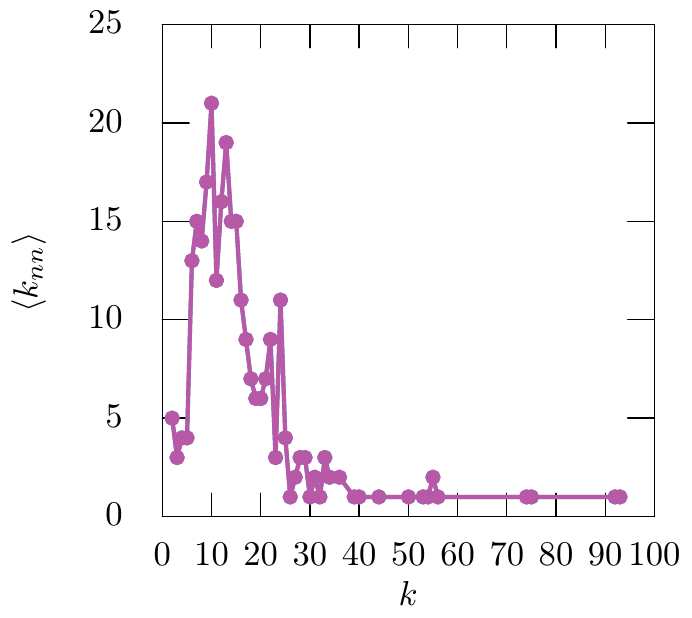}
}
\subfigure[]{
\includegraphics[width=4.5cm]{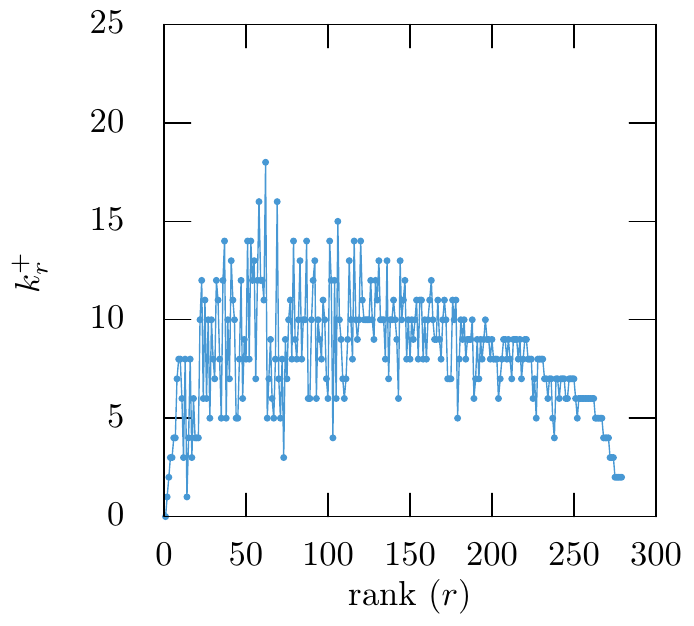}
}
\subfigure[]{
\includegraphics[width=4.5cm]{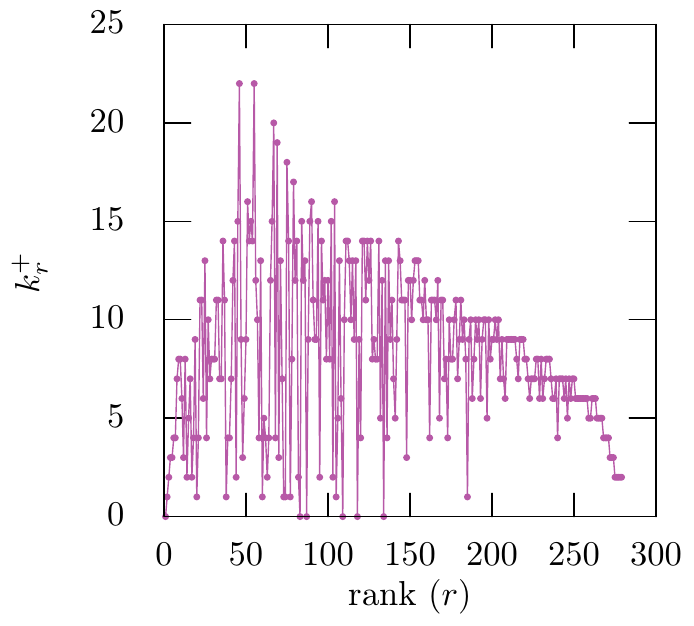}
}
\end{center}
\caption{\label{fig:assoEntro}(a) Average neighbours degree for the \emph{C.~elegans} and an ensemble with almost identical $\langle\averageDegDeg(k)\rangle$ as the original network. The percent error $\sum_i | \langle k^{(2)}_{nn}(k_i)\rangle - \langle k^{(1)}_{\rm nn}(k_i)\rangle |/N_k=3\times 10^{-4}$, where the superscript (1) refers to the original dataset and (2) to a obtained ensemble and $N_k$ is the number of different degrees present in the network. The $\{\linksRC_i\}$ sequences for (b) the original networks and (c) for the modified ME2  ensemble.}
\end{figure}

Notice the value of $\rho$  or average neighbours degree $\langle \averageDegDeg(k)\rangle$ do not define an ensemble uniquely. 
Fig.~~\ref{fig:assoEntro}(a) shows the sequence $\linksRC_i$  for the \emph{C.~elegans} and Fig.~\ref{fig:assoEntro}(b) shows the average neighbours degree for two ensembles, one obtained from the original dataset the other obtained by a modified ME2 method. The two curves are almost undistinguishable. Fig.~\ref{fig:assoEntro}(c)-(d) shows that these ensembles have different  $\{ \linksRC_i \}$ sequences. The entropy per node of \emph{C.elegans} dataset is $S=5.36$ and for the other ensemble $S=5.28$.

\subsection*{Weighted networks}
For the case where the links are weighted $\mu_i$ and the network is described by the sequences $\{\mu_i\}$ and $\{\mu^+_i\}$ the maximal entropy solution is still given by Eqs~(\ref{eq:mainResult}) and (\ref{eq:stepFunct}). For the case that $\mu_i=\degree_i$ and  $\mu^+_i$ is not restricted to be an integer the degree--degree correlation tend to be  `smoother' than when $\mu^+_i$ is restricted to the integers.  This shows that not also the structural cut--off degree introduces degree--degree correlations but also there are other correlations related to the discretisation of the links weights. (see Fig.~\ref{fig:compConfWFunction}(a))

\subsubsection*{Approximation to $m^+_i$ and the structural cut--off degree.}
If the maximal degree is less than the structural cut--off degree,  the solution of probabilities in Eq.~(\ref{eq:mainResult}) can be approximated with the configuration model where $p_{ij}=(\degree_i \degree_j)/L^2$ and $\eProb_{ii}=0$. Notice that the configuration model satisfy the conditions that $\sum_{j}p_{ij}=\degree_i/L$ and $\sum_i\sum_j \eProb_{ij}=1$ which are two of the constraints used for evaluating the maximal entropy solution.  Using this approximation to $\eProb_{ij}$ in Eq.(\ref{eq:knn}) we recover the well known result for uncorrelated networks $\averageDegDeg(k)=\langle k^2 \rangle/\langle k \rangle$~\cite{pastor2001dynamical}.

As we are interested in the connectivity within the well connected nodes, from the configuration model the number of links that node $i$ shares with nodes of higher rank is
\begin{equation}
m^{+}_i =L\sum_{n=1}^{i-1}p_{ij}=\degree_i\sum_{n=1}^{i-1}\frac{\degree_n}{L}= \frac{\degree_i}{N\langle k \rangle} \sum_{n=1}^{i-1}\degree_n.
\label{eq:approConfMod}
\end{equation} 

\begin{figure}
\begin{center}
\subfigure[]{\includegraphics[width=5cm]{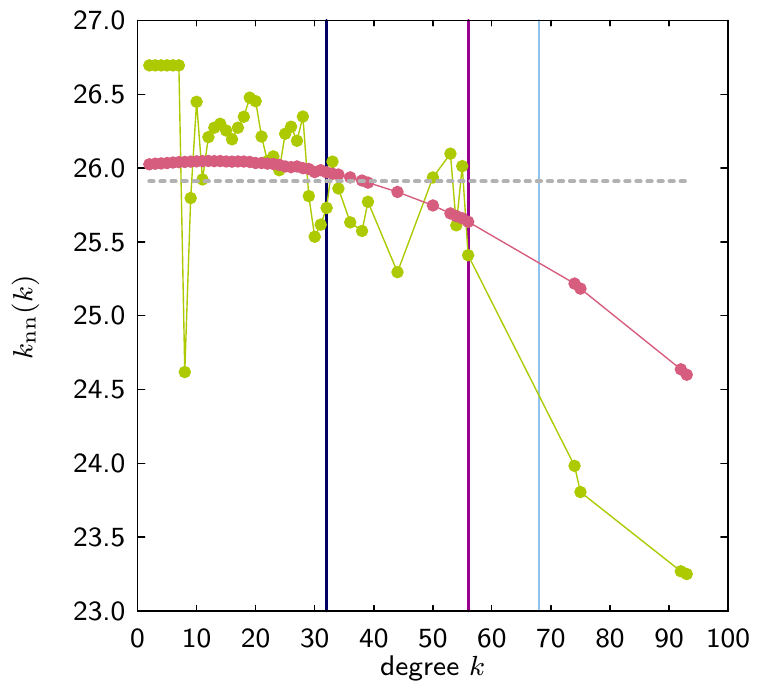}}
\subfigure[]{\includegraphics[width=5cm]{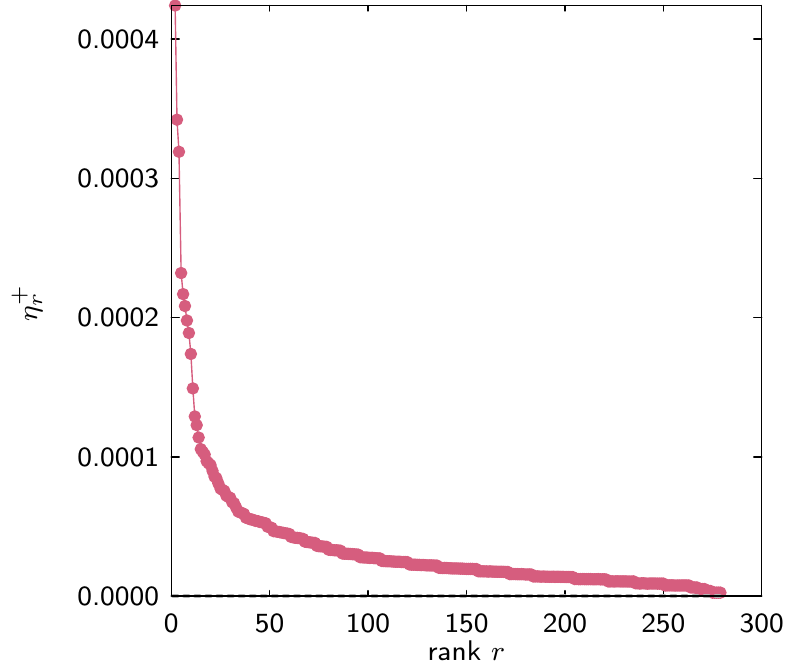}}
\end{center}
\caption{\label{fig:compConfWFunction}
\emph{C.~elegans} where the degree sequence $\{\degree_i\}$ is given and all its values are integers. (a) Value of the weights $\mu_r^+$ obtained from the ensemble ME3 (pink) and using Eq.~(\ref{eq:approConfMod}). The horizontal line is the value of $\langle k_{\rm nn}(k)\rangle$ for the decorrelated network. (b) Relative error $\eta_r^+ = (m_r^+-\mu_r^{+})/m_r^+$, where $m_r^+$ is obtained from Eq.~(\ref{eq:approConfMod}) and $\mu_r^{+}$ is obtained numerically from the ME3 ensemble. The rightmost vertical line corresponds to the structural cut--off $\sqrt{N\langle k \rangle}$, the middle vertical line is the cut--off $\degreeCF$ corresponding to the restrictions of the ME2 ensemble and the leftmost vertical line is $\degree_{\ell}$ corresponding to the restriction of the ensemble ME3.
}
\end{figure}

Figure~\ref{fig:compConfWFunction}(b) shows the relative error $\eta=1-\mu^+_i/m^{+}_i$ where $\mu_i$ was obtained numerically using ensemble ME3 for the \emph{C. elegans}.

It is also possible to obtain a better approximation to the structural cut-off degree from the configuration model, using the bound that multilinks are not allowed, i.e. $m^+_r\le r-1$,  then the bound is the largest  value  $r_{\rm max}$ where this condition 
\begin{equation}
\label{eq:structCutOff}
\frac{\degree_r}{L}\sum_{n=1}^{r-1}\degree_n > r-1
\end{equation}
still holds, the cut--off degree is $\degreeCF=\degree_{r_{\rm max}}$. 

The above structural cut--off degree assumes that the probability that a node has a self loop is small, which it would be the case multiple links between nodes are allowed. The total number of links assigned by the configuration model is given  $D = \sum_i m_i^+$ which if all the links were assigned between different nodes $D=L/2$. If this is not the case, to remove the degree--degree correlations due to the structural cut--off, the excess of links should be distributed as self loops. The probability that node $i$ has a self loop is $k_i^2/L$ then the cut--off degree is given by finding the largest value  $j=j_{\rm max}$  such that
 \begin{equation}
\label{eq:cutOffLoops}
\frac{L}{2}-\sum_{i=1}^N m_i^+ -\sum_{n=1}^j \frac{k_n^2}{L} > 0
 \end{equation}
 still holds, the cut--off degree is $\degreeCF=\degree_{j_{\rm max}}$.

Table~\ref{tab:assor} shows the cut-off degree, maximal degree and assortativity of the data and the assortativity from the null model. The ME1 produces ensembles with similar correlations as the data. The ME2 and ME3 produce decorrelated ensembles if the maximal degree is lower than the cut-off degree.
Notice that the structural cut--off based on the connectivity of the well connected is smaller than the cut--off based only in the configuration model, i.e. $\degreeCF<\degreeMax$.

\begin{figure}
\begin{center}
\includegraphics[width=4.5cm]{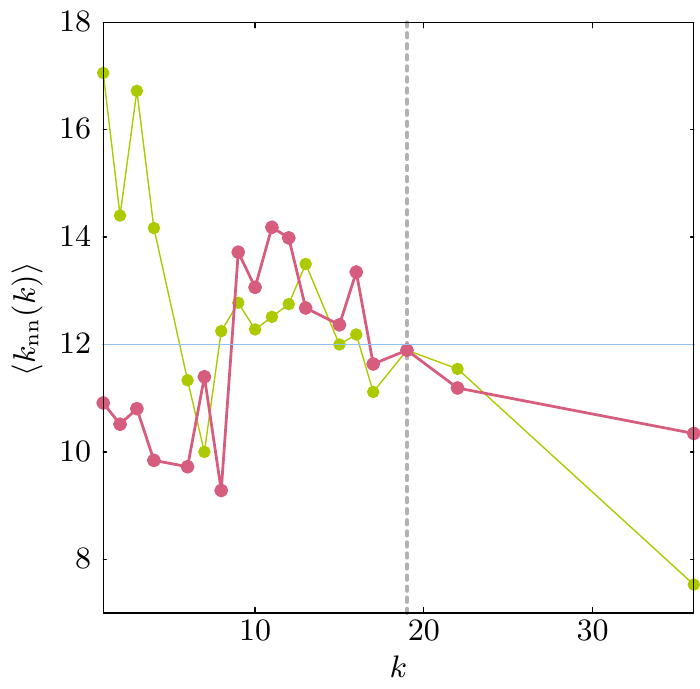}
\caption{\label{fig:lesMisCor} Original $\langle k_{\rm nn}\rangle$ for the Les-Mis network (green) and for the ME1 ensemble (pink). Notice that in this case $\rho_{\rm ME1}=0.005$ will not capture the correlations of the ensemble. The horizontal line shows the value of the decorrelated network $\langle k^2\rangle/\langle k \rangle= 12.0$ and the vertical line the structural cut--off degree ($\degreeCF=19$) obtained from Eq.~(\ref{eq:structCutOff}).
}
\end{center}
\end{figure}

Notice that for the Les-Mis network the ensemble ME1 has an assortativity coefficient of a decorrelated network. This network is an example where the assortativity coefficient  can not give a definite answer about the degree-degree correlations, see Fig.~(\ref{fig:lesMisCor}). In Les-Mis network the maximal degree is larger than the structural cut--off degree so there is a correlation due to finite size effects.
\begin{table}
\caption{\label{tab:assor}  Assortativity coefficient for the ensembles of different real networks, the maximal degree and the structural cut--off degrees. The table shows the structural cut--off $\degreeCF$ obtained from Eq.~(\ref{eq:structCutOff}), $\sqrt{N\langle k \rangle}$ and from Eq.~(\ref{eq:cutOffLoops}). Their values are only shown if they are smaller than the maximal degree $k_{\rm max}$.}
\begin{center}
\begin{tabular}{l | c | c | c | c | c | c | c |c|}
Network & $\rho_d$ & $\rho_{ME1}$ & $\rho_{ME2}$ & $\rho_{ME3}$ & $\degreeMax$ & $\degree_{r_{\rm max}}$ & $ \sqrt{N\langle k \rangle}$ & $\degree_{j_{\rm max}}$\\ \hline
Adj nouns & -0.129 & -0.125 & -0.085 & -0.047 & 49 & 28 & 29.15 & 15\\
Airports & -0.267& -0.223 & -0.264 & -0.017 & 145 & 64 & 77.20 & 91\\
Astro & 0.235 & 0.254 & -0.002 & 0.000 & 360 & - & - & -\\
\emph{C.~elegans} & -0.091 & -0.035 & -0.030 & -0.017 & 93 & 56 & 67.68 & 32\\
Dolphins & -0.043 & -0.027 & -0.050 & -0.045 & 12 & - & - & -\\
Football & 0.162 & 0.136 & -0.024 & -0.005 & 12 & - &- & -\\
Hep-Th & 0.293 & 0.321 & -0.012 & 0.032 & 50 & - & - & -\\
AS-Internet & -0.194 & -0.188 & -0.176 & -0.042 & 2389 & 116 & 216.37 & 1334\\
Karate & -0.475 & -0.434 & -0.205 & -0.114 & 17 & 12 & 12.49 & 12 \\
Net Sci & -0.081 & -0.025 & -0.018 & -0.010 & 34 & - & - & -\\
Political blogs & -0.221 & -0.153 & -0.046 & -0.007 & 351 & 149 & 182.86 & 116\\
Political books & -0.127 & -0.135 & -0.021 & -0.018 & 25 & - &- & -\\
Power & 0.003 & 0.035 & -0.022 & 0.030 & 19 & - & - & -\\
Protein & -0.136 & -0.080 & -0.007 & -0.005 & 282 & 147 & 172.31 & 46\\
Random ER & -0.004 & -0.002 & -0.012 & 0.035 & 13 & - & - & -\\
Les Mis & -0.165 & 0.005 & -0.079 & -0.065 & 36 & 19 & 22.54 & 15\\
\end{tabular}
\end{center}
\end{table}

\subsection*{Comment about the rich-club}
Notice that the ranked based rich--club coefficient~\cite{Zhou04} is $\richClubRank_r = (2/(r(r-1))) \sum_{i=1}^r \linksRC_i$, thus conserving $\linksRC_i$ is equivalent to the conservation of the rich--club coefficient. 

The weighted ME3 model can be used to evaluate the  normalised rich--club~\cite{Colizza06}. The uncorrelated rich--club coefficient is
\begin{equation}
\phi_{\rm unc}(r) =
\frac{2}{r(r-1)N\langle k \rangle}\sum_{i=2}^{r}\degree_i\left( \sum_{n=1}^{i-1}\degree_n\right)
\end{equation}
From Eq.~(\ref{eq:approConfMod}) the normalised weighted rich-club is
\begin{equation}
\phi_{\rm norm}(r) = \frac{N\langle k \rangle \sum_{i=1}^r \linksRC_i}{\sum_{i=2}^r\degree_i\left(\sum_{n=1}^{i-1}\degree_n\right)}.
\end{equation}

\subsubsection*{Clustering coefficient from the ensemble and realisation of the networks}
The local clustering coefficient of node $i$ is the number of triangles $t_i$ that contain node $i$ normalised by the number of possible triangles that node $i$ can have 
\begin{equation}
C_i(\degree_i)=\frac{t_1}{\degree_i(\degree_i-1)/2}.
\end{equation}
In our case the probability that there is a triangle between nodes $i$, $j$ and $k$ is $P(ijk)=\eProb_{ij}\eProb_{jk}\eProb_{ki}$ and the average number of triangles between these three nodes is $\langle t_{ijk}\rangle = L^3\eProb_{ij}\eProb_{jk}\eProb_{ki}$. If the network is uncorrelated then we can use the configuration model and $\langle t_{ijk}\rangle = \degree_i^2\degree_j^2\degree_k^2/L^3$.  For networks that only have degree--degree correlations the distribution $\eProb_{ij}$ determines the distribution of triangles and hence the clustering coefficient.

Notice that the ensembles are constructed using soft constraints, that is the constraint is on the average, that means that it is possible to have low ranking nodes (average small degree) that have many triangles. The extreme case is nodes with average degree one that nevertheless on average  can be members of a triangle. This is because the chance that this kind of node has more than one link is not negligible. 

\subsection*{Information gain using different ensembles}
The information gain is measured via the Kullback--Leibler divergence, in this case it is used to measure how much information is gained if the network is described using ME1 instead of ME2 or ME3. To compare the change form ensemble ME2 to ME1 then
\begin{equation}
\label{eq:infoLoss}
D\left(p^{[ME1]}||p^{[ME2]}\right) = \sum_{i=1}^N\sum_{j=1,j\ne i}^N p^{[ME1]}_{ij} \log \left(\frac{p^{[ME1]}_{ij}}{p^{[ME2]}_{ij}} \right)
\end{equation}
where $p^{[ME2]}_{ij} \ne 0$ for all $i$ and $j$. This last condition is satisfied by the de--correlated ensemble. Notice that in the correlated ensemble ME1 it is possible to have $p_{ij}^{[ME1]}=0$, for example when $i<j$ and $\linksRC_j=0$ (see Eq.~(\ref{eq:mainResult})), in this case we assume that  $x\log(x)= 0$ if $x= 0$.  Table~\ref{tab:KL} shows the information gain for some real networks. The information decreases as the restrictions on the sequence $\{\linksRC_i\}$ are relaxed.

\begin{table}
\caption{\label{tab:KL}
Information gain comparing the  probabilities obtained from the ensembles.
}

\begin{center}
\begin{tabular}{l | c | c | c}\\
\hline
Network & $D\left(p^{[ME1]}||p^{[ME2]}\right)$ & $D\left(p^{[ME1]}||p^{[ME3]}\right)$ & $D\left(p^{[ME2]}||p^{[ME3]}\right)$\\ \hline
Adj-nouns & 0.082 & 0.094& 0.010\\
Astro & 0.330& 0.332&0.002 \\
C. elegans & 0.084 & 0.085& 0.003 \\
Airports & 0.100 & 0.157 & 0.057\\
Dolphins & 0.191 & 0.202 & 0.014 \\
Net. Scientists & 0.235 & 0.243 & 0.029\\
Football & 0.106 & 0.100& 0.009\\
Hep-Th & 0.351& 0.335& 0.006\\
AS-Internet & 0.243&  0.366& 0.109\\
Karate & 0.213& 0.232& 0.032\\
Les Mis & 0.184 & 0.183 & 0.011 \\
Pol. books & 0.128& 0.133& 0.011\\
Power & 0.311 & 0.357 & 0.069 \\
Protein & 0.148 & 0.156& 0.013\\
Random Network & 0.206 & 0.224 & 0.032\\
Pol. blogs & 0.077 & 0.083 & 0.003\\
\end{tabular}
\end{center}
\end{table}

\subsection*{Networks generation from the ensemble}

\begin{figure}
\begin{center}
\subfigure[]{\includegraphics[width=4.5cm]{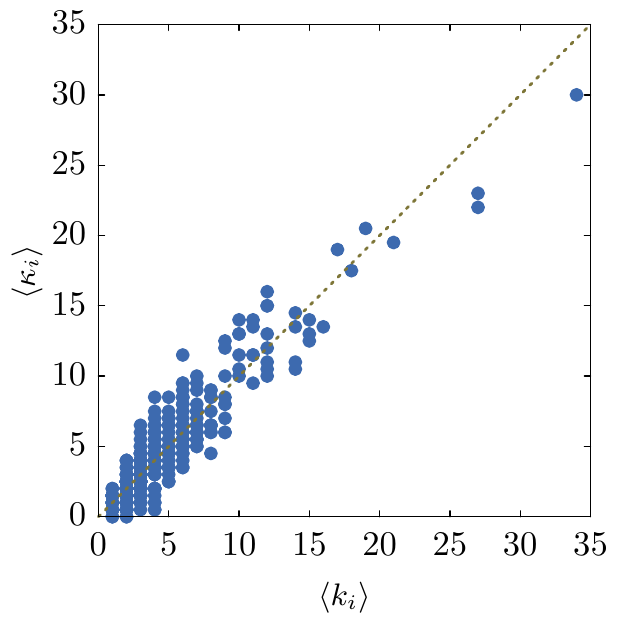}}
\subfigure[]{\includegraphics[width=4.5cm]{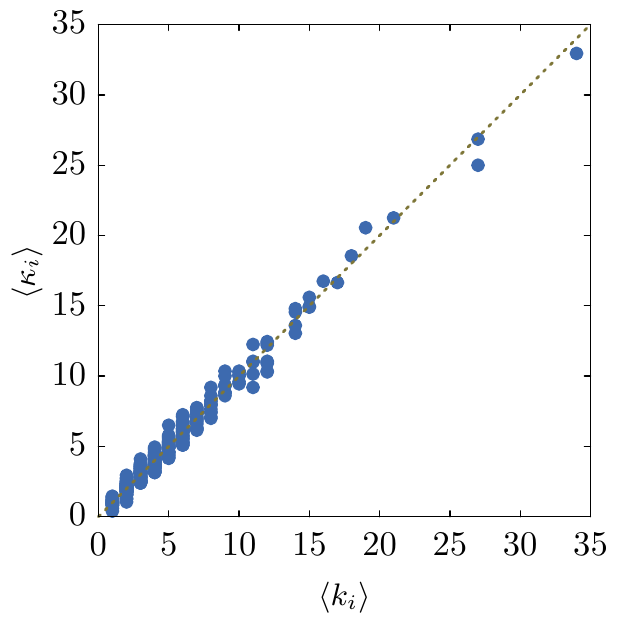}}
\subfigure[]{\includegraphics[width=4.5cm]{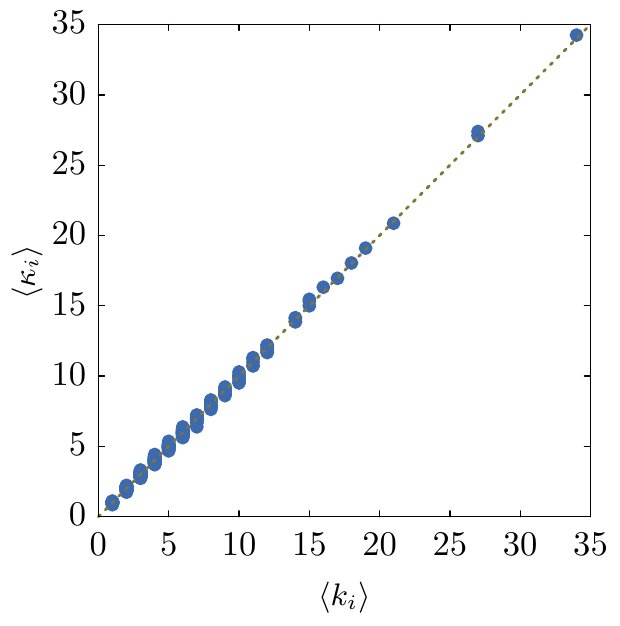}}
\end{center}
\caption{\label{fig:approSoft}
Average degree evaluated from several network realisations vs. the degree of the \emph{C. elegans} network. (a) 10 realisations, (b) 100 realisations and (c) 1000 realisations.}
\end{figure}
To generate a network that is a member of the ensemble we use a Bernoulli process where
the existence of a link between nodes $i$ and $j$ is given by $p_{ij}$. The process is carried out until there are $L$ links in the generated network.  Fig.~\ref{fig:approSoft}(a) shows the average degree $\langle \degree_i \rangle_m$  obtained from $m$ realisations of the ensemble when multiple links are allowed..

\subsection*{Restricted randomisation}
The other procedure to generate a network ensemble is restricted randomisation~\cite{maslov2002specificity}, where the degree sequence is always fixed and the $\{\linksRC_i\}$ can be fixed or not. As in the ensembles generated via the maximal entropy method, the conservation of the sequences $\{\degree_i\}$ and $\{\linksRC_i\}$ generates networks with similar degree--degree correlations as the original network. Table~\ref{tab:assoRestRand} compares the assortativity coefficient $\rho$ for some real networks and the average $\langle \rho_{\rm res}\rangle$ obtained from the restricted randomisations. 

\begin{table}[h]	
\begin{center}
\caption{\label{tab:assoRestRand} Assortativity coefficient for different networks obtained by the restricted randomisation which conserves the sequences $\{\degree_i\}$ and $\{\linksRC_i\}$. The assortativity coefficient $\rho_{\rm res}$ were obtained by switching links $1000\times L$ times. 
}
\begin{tabular}{c c c }
Network & $\rho$ & $\langle \rho_{\rm res}\rangle $ \\ \hline
Adj nouns & -0.129 & $-0.199\pm 0.014$  \\
Airports &-0.267 &$-0.283\pm 0.001$ \\
Protein & -0.136& $-0.118 \pm 0.001$\\
Random & -0.045 & $-0.116 \pm 0.004$\\
\emph{C. elegans} &-0.092& $-0.094\pm 0.005$\\
NetSci & -0.081 & $-0.101\pm 0.011$\\
AS-Internet & -0.194& $-0.195 \pm 0.000$\\
Karate & -0.475 & $-0.457 \pm 0.018$ \\
LesMis & -0.165 & $-0.098\pm 0.022$ \\
PolBooks & -0.127& $-0.177\pm 0.012$\\
PolBlogs & -0.221& $-0.219 \pm 0.002$ \\ \hline
Astro & 0.235& $0.154 \pm0.001$ \\
Football & 0.162& $0.080 \pm 0.012 $ \\
HepTh & 0.185& $0.069 \pm 0.004$\\
Power & 0.003& $-0.060\pm 0.005$\\
\end{tabular}
\end{center}
\end{table}

\begin{figure}[h]
\begin{center}
\subfigure[]{\includegraphics[width=4.5cm]{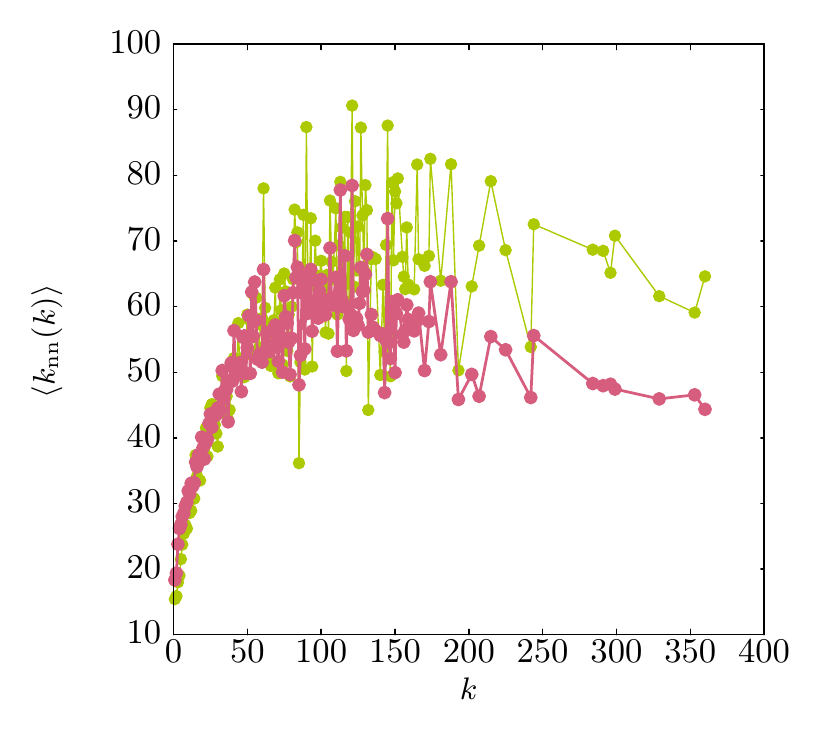}}
\subfigure[]{\includegraphics[width=4.5cm]{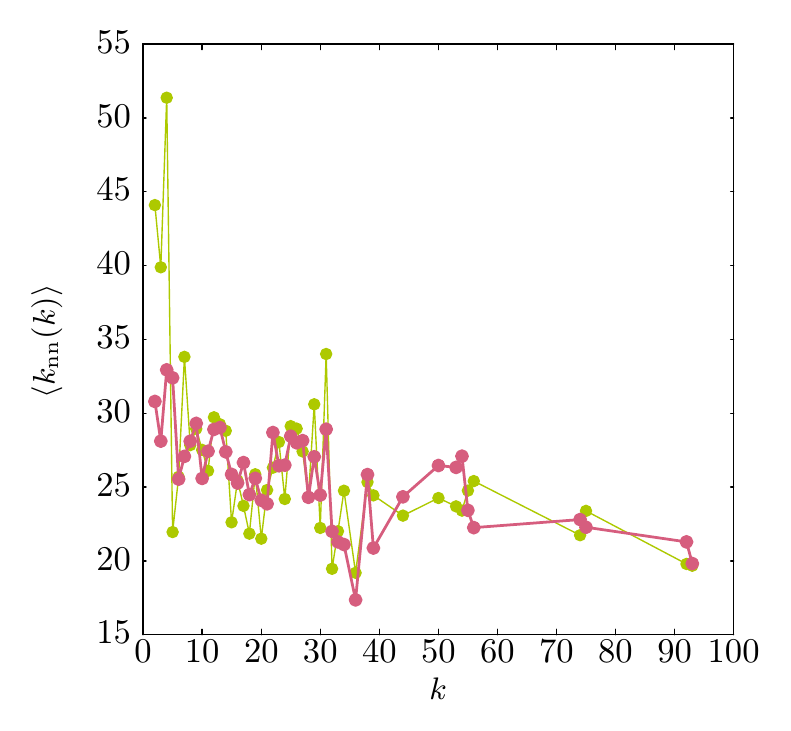}}
\subfigure[]{\includegraphics[width=4.5cm]{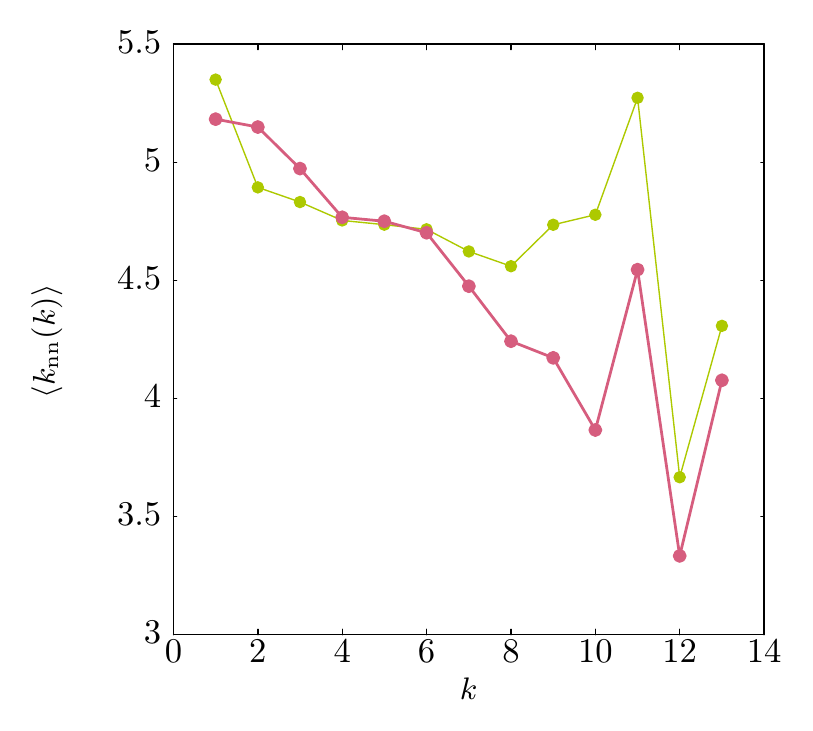}}
\caption{\label{fig:propRestRand}Comparison of the average neighbours degree for the original network and the average obtained from the restricted randomisation. (a) The Astrophysics-collaborators, which is disassortative. (b)The \emph{C. elegans} which is disassortative. (c) An a random network, which has a spurious correlation that cannot be removed by the restricted randomisation as the $\{\linksRC_i\}$ sequence is conserved.}
\end{center}
\end{figure}

Fig.~\ref{fig:propRestRand} shows the average neighbours degree $\langle \averageDegDeg \rangle$ for several real networks, confirming that conserving the sequences $\{\degree_i\}$ and $\{\linksRC_i\}$ generates networks with similar degree--degree correlations. Notice that for a random network the randomisation does not generates a de-correlated network as the correlation that was present in the original network cannot be removed.

%\newpage
%\bibliographystyle{abbrv}
%\bibliography{BeyondRich.bib}
%\putbib
%\end{bibunit}
\newpage
\newpage
%\begin{bibunit}
\section*{Appendix Cores}

\subsection*{Spectral cores}
We assume that the nodes are ranked in decreasing order of their degree and that the networks connectivity is described by the degree sequence $\{k_i\}$ and the sequence $\{\linksRC_i\}$.  If ${\bf A}$ is the adjacency matrix where $A_{ij}=A_{ji}=1$ if nodes $i$ and $j$ share a link and zero otherwise. The spectrum of the graph is the set of eigenvalues $\Lambda_1\ge \Lambda_2\ge \ldots \ge \Lambda_N$ of the matrix ${\bf A}$ where $\Lambda_1$ is the spectral radius. 

A lower bound for $\Lambda_1$ is~\cite{van2010influence,van2010graph} %need an older reference
$\Lambda_1 \ge \left( \walk_{2n}/\walk_0\right)^{1/(2n)}\ge  \left( \walk_{n}/\walk_0\right)^{1/n}$, $n=1,\ldots ,$
where $\walk_n= \overline{u}^T{\bf A}^n \overline{u}$ is the total number of walks of length $n$, ${\bf A}$ is the adjacency matrix and $\overline{u}$ is a vector with all its entries equal to one. An upper bound for the number of walks is $\walk_n\le \sum_{j=1}^{N}\degree_j^n$~\cite{fiol2009number} where the equality is true only if $n\le 2$. 
The idea behind the bounds based on the hubs is to evaluate the density of walks of length one (or two) that include at least one of the hub nodes. 

\subsection*{Bound based on walks of length one}
Using the bound for $\walk_n$ for $n=1$ we define
\begin{eqnarray}
\nonumber
g(r)&=& \frac{1}{r}\sum_{i=1}^r \degree_i
= \frac{1}{ r}\sum_{i=1}^r \left( 2\linksRC_i + \degree_i -2\linksRC_i\right)
= \frac{1}{r}\sum_{i=1}^r  2\linksRC_i  + \frac{1}{r}\sum_{i=1}^r \left( \degree_i -2\linksRC_i\right)\\
\nonumber
&=&2 \langle \linksRC\rangle_r +  \langle \degree-2\linksRC\rangle_r
,\quad r=1,\ldots , N.
\end{eqnarray}
The sum containing only terms of the form $2\linksRC_i$ gives the average number of links within the top $i$ ranked nodes. The other sum containing the terms $\degree_i-2\linksRC_i$ is the average number of links between the top $i$ ranked nodes and nodes of lower rank. Notice that  if $r=N$ then $2 \langle \linksRC\rangle_N =2L/N$, $\langle \degree-2\linksRC\rangle_N=0$ and $g(N)=2L/N$,  which is the well known lower bound $B_1=\walk_1/\walk_0 = 2L/N\le \Lambda_1$. Also notice that $2 \langle \linksRC\rangle_r$  could be larger than $g(N)=2L/N$. We split the network into two parts by considering  the value $r$ such that $2 \langle \linksRC\rangle_r$ is maximal, that is when the density of connections between the top ranked nodes is maximal. In this case the core of the network is the nodes of rank greater than $r_c$  where
\begin{equation}
\label{eq:defRc}
r_c^{(1)} = \underset{r}{\max}\left( \left\{ \underset{r}{\mathrm{arg~max}} \left(2 \langle \linksRC\rangle_r\right)\right\} \right)
\end{equation}
where the superscript in $r_c^{(1)}$ is used to label this bound. The bound is~\cite{mondragon2016network}
\begin{equation}
b_1 = 2\langle \linksRC \rangle_{r_c^{(1)}}\le \firstEigenvalue.
\label{eq:linearBound}
\end{equation}
Notice that if $r^{(1)}_c=N$ then $b_1=\walk_1/\walk_0$ which is a well know bound of $\firstEigenvalue$.

\subsection*{Bound based on walks of length two}
The above bound can be improved if we consider the connectivity of the well connected nodes and the connectivity of their neighbouring nodes.  In this case we consider walks of length two $\walk_2$.  The number of walks of length two starting from node $j$, $\walk_2(j)$, is the same as the walks of length one starting from the neighbouring nodes of $j$, we denote the neighbours of $j$ as $j_q$. Then
\begin{equation}
\label{eq:walkTwo}
\walk_2(j) = \sum_{q=1}^{k_j}\walk_1(j_q).
\end{equation}
If we distinguish which walks of length 1 end on one of the top $r$ ranked nodes then
\begin{equation}
\label{eq:walkTwoAppro}
\walk_2(j) = \sum_{q=1}^{k_j}\left(\walk_1(j_q)\Theta(j_q,r)+\walk_1(j_q)(1	-\Theta(j_q,r))\right)
\end{equation}
where $\Theta(j_q,r)$ is the step function $\Theta(a,b) = 1$ if $a<b$ and zero otherwise. We are interested in the first term, $\sum_{j_q=1}^{k_i}\walk_1(j_q)\Theta(j_q,r)$, which is the number of links that the nearest neighbours of $j$ have a link with a node with rank equal of less than $r$, we denote this degree with $K^{(1)}_j(r)$. Similarly as the bound $b_1$, we evaluate the density of  these walks using 
\begin{equation}
h(r) = \frac{1}{r}\sum_{j=1}^r K^{(1)}_j(r)
\end{equation}
if 
\begin{equation}
r_c^{(2)} = \underset{r}{\max}\left( \left\{ \underset{r}{\mathrm{arg~max}} \left(  h(r)\right)\right\} \right)
\end{equation}
  the bound is 
\begin{equation}
b_2=\sqrt{h(r_c^{(2)})} \le \firstEigenvalue
\end{equation}
Notice that if $r^{(2)}_c=N$ then $b_2=\walk_2/\walk_0$, which is the well know bound $\walk_2/\walk_0\le \firstEigenvalue^2$.

For comparison purposes we compared these two lower bounds with the lower bounds $B_1= \walk_1/\walk_0$, $B_2= \sqrt{\walk_2/\walk_0}$ and~\cite{van2010influence}
\begin{equation}
B_{3}=\left( \frac{\walk_3}{\walk_0} \right) ^{1/3}= \left( \frac{1}{\walk_0}\left( \rho \left( \sum_{i=1}^N \degree_i^3-\frac{\walk_2^2}{\walk_1}\right)+\frac{\walk_2^2}{\walk_1}\right)\right)^{1/3}
\end{equation}
where this last bound can be expressed  as a function of the assortativity coefficient $\rho$~\cite{Newman02,van2010influence}.
Also we consider the optimised bound~\cite{van2010influence} based on walks of length one, two and three, 
\begin{equation}
B_M = \frac{\walk_0\walk_3 - \walk_1\walk_2 + R}{2(\walk_0\walk_2-\walk_1^2)}
\end{equation}
where 
\begin{equation}
R=\sqrt{\walk_0^2\walk_3^2-6\walk_0\walk_1\walk_2\walk_3-3\walk_1^2\walk_2^2+4(\walk_1^3\walk_3+\walk_0\walk_2^3)}
\end{equation}
Notice that by construction $b_1\ge B_1$ and $b_2 \ge B_2$. 
Table~\ref{tab:bounds} compares the bounds of $\firstEigenvalue$ for different real networks. The bound $B_M$ gives the best approximation of the $\firstEigenvalue$ except for the Hep-Th, Power and the AS-Internet networks. However the bounds $b_1$ and $b_2$ are simple to evaluate and have a simple interpretation in terms of the connectivity of the network.

\begin{table}
\begin{center}
\begin{tabular}{l r r r  r | r r | r | r r r}
Network & $B_1$ & $B_2$ & $B_{3}$ & $B_M$ & $b_1$ & $b_2$ & $\firstEigenvalue$ & $r_c^{(1)}$ & $r_c^{(2)}$ & $N$ \\ \hline
Nouns & 7.58 &  10.22& 10.94 & 12.66 & 9.30& { 11.34}& 13.15& 49& 66& 112\\
Airports & 11.92 & 25.32 & 30.49 & 44.77 & {{ 38.02}} & { 42.70} & 48.07 & 71& 71& 500\\
C. elegans & 16.40 & 20.62 & { 21.82}& 24.58 & 17.99 & 21.22 & 25.94 & 172 & 197 & 279\\
Dolphins & 5.12 & 5.90 & 6.17 & 6.75 & 6.04 & { 6.46} & 7.19 & 41 & 39 & 62\\
Football & 10.66 & 10.69 & { 10.71} & 10.74 & 10.66 &10.69 &10.78 &115 & 115 &115\\
Hep-Th & 4.13 & 5.99 &7.25 & 11.00 &{{ 9.82}}  & { 15.05} & 23.00 & 70 & 70 & 7610\\
Karate & 4.50 & 5.97 & { 5.98} &6.50 & 5.00 & { 5.98} & 6.72 & 22 & 33 & 34 \\
Les Miss &  6.59 & 8.91 & 9.59 & 11.20 &{{ 10.00} }& { 10.97} & 12.00&  28 & 24 & 77\\
Net Sci & 4.82 & 6.21 & 6.64 & 7.62 & 5.69 & { 7.19} & 10.37 & 192 & 4  & 379\\
Political blog & 27.31 & 47.11 & 53.21 & 69.08 & {{54.45}} & { 63.16} & 74.08 & 323 & 321 &1224\\
Political book & 8.40 & 10.01 & { 10.46} & 11.48 & 8.85 & 10.33 & 11.93 & 68 & 68 & 105 \\
Power & 2.66 & 3.21 & 3.42 & 3.87 & 2.88 & { 4.44} & 7.48 & 3715  & 32 & 4941\\
Protein & 6.30 & 12.34 & 13.54 & 17.38 & 12.86 & { 16.79} & 21.16 & 733 & 1 & 4713\\
AS-Internet & 4.18 & 33.35 & 28.51 &41.81 & 20.20 & { 48.97} & 60.32 & 77 & 2 & 11174\\
& \\
\end{tabular}
\caption{\label{tab:bounds} The spectral radius $\firstEigenvalue$ and its bounds $B_1$, $B_2$ and $B_3$ based on the sum of all walks of length one, two and three respectively.  Bounds $b_1$ and $b_2$ obtained from local walks from the core, or from the local connectivity of the core nodes. The entries $r_c^{(1)}$ and $r_c^{(2)}$ are the number of nodes that constitute the core obtained from $b_1$ and $b_2$. The entry $N$ is the total number of nodes in the network. 
}
\end{center}
\end{table}

\begin{figure}
\begin{center}
\subfigure[]{\includegraphics[height=4.5cm]{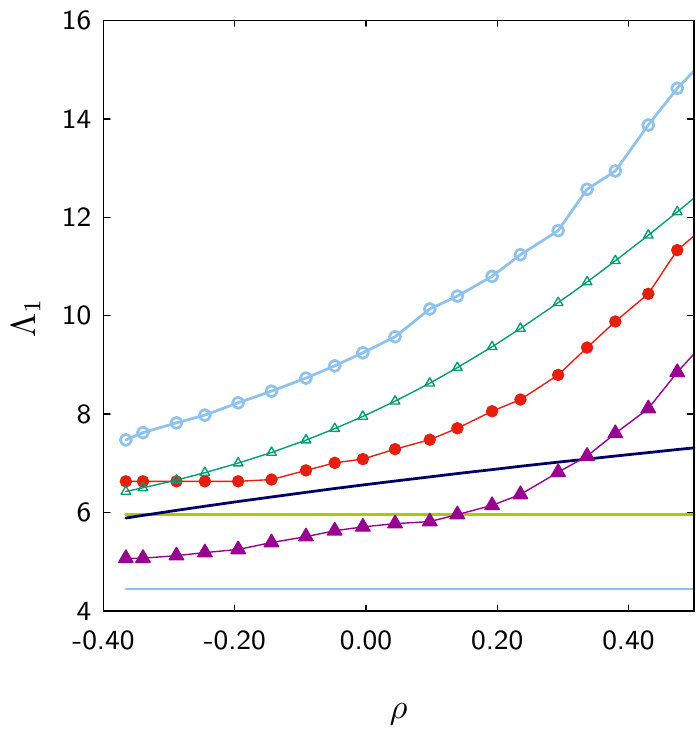}}
\subfigure[]{\includegraphics[height=4.5cm]{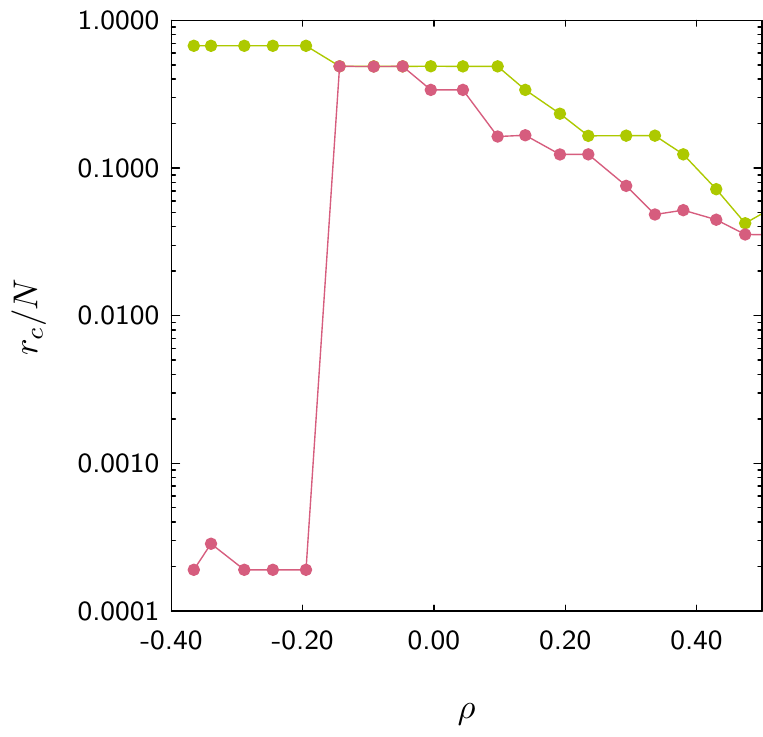}}
\subfigure[]{\includegraphics[height=4.5cm]{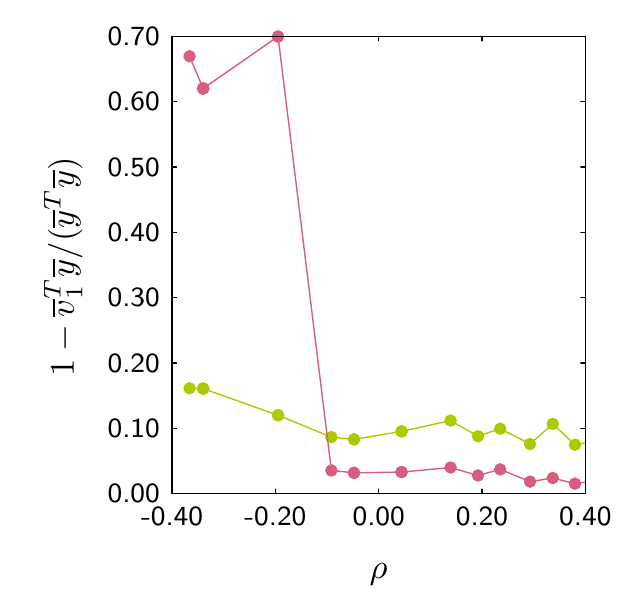}}
\end{center}
\caption{\label{fig:spectraRhoDep}
(a) Dependance of the different bounds as a function of the assortativity coefficient. (b) Size of the core as a function of the assortativity coefficient. (c) Relative error between the eigenvector $\overline{v}$ and its approximation $\overline{y}$.
}
\end{figure}

Figure~\ref{fig:spectraRhoDep}(a) shows the behaviour of the bounds as a function of the assortativity coefficient. It seems that the bound $b_2$ can produce better bounds that $B_M$ for networks with high assortativity or disassortativity coefficient, perhaps this is the reason that this bound is better for the Hep-Th, Power and AS-Internet networks than the $B_M$ bound. Figure~\ref{fig:spectraRhoDep}(b) shows the size of the core obtained from the bound $b_1$ (green) and $b_2$ (pink). Notice the drastic change on the core size obtained from the bound $b_2$ when the network becomes more disassortative. 

To confirm that $b_1=2\langle \linksRC \rangle_{r_c}$ is a bound of the spectral radius consider Rayleigh's inequality $\Lambda_1\ge \overline{u}^T{\bf A}\overline{u}/(\overline{u}^T\overline{u})$.
If ${\bf A}$ is the adjacency matrix of a network ranked in decreasing order of its node's degree and  $\overline{u}$ is a vector with 1 in the top $r_c$ entries and 0 otherwise then 
\begin{equation}
\label{eq:degreeVect}
\overline{y} = {\bf A}\overline{u}=
\left(
\begin{matrix}
\begin{array}{c c c | c c c c}
A_{1,1}& \hdots& A_{1,{r_c}} &A_{1,{r_{c}+1}} & \hdots & &A_{1,N}\\
\vdots& \ddots& & & \ddots& &\\
& & & & & &\\
A_{r_c,1}&\hdots & & & & &\\ \hline
A_{{r_{c}+1},1}& \hdots& & & & &\\
\vdots &  \ddots& & & \ddots & &\\
& & & & & &\\
A_{N,1}&\hdots & & & & &\\
\end{array}
\end{matrix}
\right)
\left(
\begin{matrix}
1\\
1\\
\vdots\\
1\\
0\\
\vdots\\
0\\
\end{matrix}
\right)
=
\left(
\begin{matrix}
K_1^+(r_c)\\
K_2^+(r_c)\\
\vdots\\
K_{r_c}^+(r_c)\\
K_{r_c+1}^+(r_c)\\
\vdots\\
K_{N}^+(r_c)\\
\end{matrix}
\right)
\end{equation}
where $K_i^+(r_c)$ is the number of links that node $i$ shares with the $r_c$ top ranked nodes. Then $\overline{u}^T{\bf A}\overline{u}=\sum_{i=1}^{r_c}K_i^+(r_c)$ is the total number of links are shared by the top $r_c$ ranked nodes, recalling that $\linksRC_i$ is the number of links between node $i$ shares nodes of largest rank then $\sum_{i=1}^{r_c}K_i^+(r_c)=2\sum_{i=1}^{r_c}\linksRC_i$. As $\overline{u}^T\overline{u}=r_c$, Rayleigh's inequality gives $\Lambda_1\ge 2\langle \linksRC \rangle_{r_c}$.
 
 For the bound $b_2$, or  $h(r_c^{(2)}) \le \firstEigenvalue^2$, the procedure is similar as for the $b_1$ case. In this case $\firstEigenvalue^2 \ge \overline{u}^T{\bf A}^2\overline{u}/(\overline{u}^T\overline{u})$.  The entries $({\bf A}^2)_{ij}$ correspond to the number of walks of length two that start in $i$ and end in $j$. If $\overline{u}$ is a vector with 1 in the top $r_c$ entries and 0 otherwise then, $\overline{u}^T{\bf A}^2\overline{u}$ is the number of walks of length two, $\walk_2(r_c,r_c)$, that start in one of the top $r_c$ nodes and end in on of these top $r_c$ nodes. Notice that $\walk_2(r_c,r_c)$  is equal to the number of links in the whole network that connect with at least one node in $r_c$, that is $\walk_2(r_c,r_c)=\sum_{i=1}^N K_i^+(r_c)$.

To measure how well the vector $\overline{y}$ approximates the eigenvector $\overline{v}_1$ we evaluate
\begin{equation}
\delta =1-\frac{\overline{v}_1^T\overline{y}}{\overline{y}^T\overline{y}}
\end{equation}
where the closer this quantity is to zero, the better $\overline{y}$ approximates $\overline{v}_1$ (see Fig.~\ref{fig:spectraRhoDep}(c)).
Notice that for highly disassortative networks the size of the spectral core obtained from the bound $b_2$ can be very small. However this is not translated into a better approximation to the eigenvector $\overline{v}_1$ but the opposite, the approximation becomes poor.

\subsection*{Relationship with the eigenvector centrality}
If the eigenvectors of the matrix ${\bf A}$ are $\{\firstEigenvector = \firstEigenvector_1,\ldots ,\firstEigenvector_N\}$  and corresponding eigenvalues $\{\firstEigenvalue,\ldots, \Lambda_N\}$ where $\Lambda_i\ge \Lambda_{i+1}$, then
$\overline{z}={\bf A}\overline{u} = c_1\firstEigenvalue (\firstEigenvector_1+(c_2\Lambda_2)/(c_1\firstEigenvalue)\firstEigenvector_2+\ldots)$ where $c_i$ are constants, then to first order approximation $\overline{y}\approx c_1\firstEigenvalue \firstEigenvector_1$, hence $\overline{y}$ is an approximation to the eigenvector centrality $\firstEigenvector$. The entries of $\overline{y}={\bf A}\overline{u}$ are $y_i=K^+_i(r_c)$, where $K^+_i(r_c)$ is the number of links that node $i$ shares with the $r_c$ top ranked nodes.
For the case of the bound considering walks of length two described by the matrix ${\bf A}^2$ the approximation to the eigencentrality is given by the vector $\overline{y}={\bf A}^2\overline{u}$ with entries $y_i=\walk_2(r_c,i)$ where $\walk_2(r_c,i)$ are the number of walks of length 2 that start in $r_c$ and end in any node $i$.

\section*{Biased random walks}

\subsection*{Maximal Rate Entropy Random walk (MERW)}
In a finite, undirected, not bipartite and connected network a random walker would jump from node $i$ to a neighbouring node $j$ with a probability $\jumpProb$. The probability that the walker is in node $j$ at time $t+1$ is $p_j(t+1) = \sum_i ({\bf A})_{ij}\jumpProb p_i(t)$.
 The probability of finding the walker in node $i$ as time tends to infinity is given by the stationary distribution $\overline{p}^*=\{p^*_i\}$.   
 In a network, the jump probability $\jumpProb$ can be expressed as
\begin{equation}
\jumpProb = \frac{({\bf A})_{ij}f_j}{\sum_j ({\bf A})_{ij}f_j}
\label{eq:defJump}
\end{equation} 
where $({\bf A})_{ij}$ is the $ij$ entry of the adjacency matrix and $f_j$ is a function of one or several topological properties of the network, in this case the stationary distribution is~\cite{gomez2008entropy}
\begin{equation}
p^*_i=\frac{f_i\sum_j ({\bf A})_{ij}f_j}{\sum_n f_n\sum_j ({\bf A})_{nj}f_j}.
\label{eq:statProb}
\end{equation}

The measure which tell us the minimum amount of information needed to describe the stochastic walk in the network is the entropy rate $s=\lim_{t\rightarrow\infty}S_t/t$, where $S_t$ is the Shannon entropy of all walks of length $t$,  which is
\begin{equation}
s=-\sum_{i=1}^N
p^*_i\sum_{j=1}^N\jumpProb  \ln (\jumpProb).
\label{eq:entrop}
\end{equation}

The maximal rate entropy $s_{\rm max}$ corresponds to random walks where all the walks of the same length have equal probability.
The value of $s_{\rm max}$ can be expressed in terms of the spectral properties of the network as
\begin{equation}
s_{\rm max}= \lim_{t\rightarrow\infty}\frac{\ln \sum_{ij}({\bf A}^t)_{ij} }{t} =\lim_{t\rightarrow\infty}\left(\frac{1}{t}\right) \ln\left(a_1^2\eigenval_1^t(1+(a_2/a_1)^2(\eigenval_2/\eigenval_1)^t+\ldots )\right)=
 \ln (\firstEigenvalue).
\end{equation}

For the MERW the probability $\jumpProb$ is such that all the walks of the same length have equal probability. 
The stationary probability for the MERW is $p^*_i=v_i^2$ where $v_i$ is the $i$ entry of the eigencentrality.

\subsection*{Core-biased random walk}
If the largest eigenvalue-eigenvector pair is not known, the MERW results suggests that a good approximation to the largest eigenvector $\firstEigenvector$ could be used to construct a biased random walk.  
A bound for the largest eigenvalue in terms of the connectivity of the nodes of high degree is $b_1=1/r\sum_{i=1}^r \linksRC_i \le \firstEigenvalue$ and an approximation to the corresponding eigenvector is  $\overline{z}(r)={\bf A}\overline{u}(r)$, where $\overline{u}$ is a vector with its top $r$ entries equal to one and the rest to zero (see Appendix: The spectral-core). The vector $\overline{z}(r)={\bf A}\overline{u}(r)$ has entries $z_i(r)=K^+_i(r)$, where $K^+_i(r)$ is the number of links that node $i$ shares with the top $r$ ranked nodes.

This bound based on $\{\linksRC_i\}$ suggests a core biased random walk. If the top $r$ ranked nodes are the core of the network, then a core-biased random jump is~\cite{mondragon2018}
\begin{equation}
P_{i\rightarrow j}(r) =  \frac{({\bf A})_{ij}(K^+_j(r)+1)}{\sum_j^N ({\bf A})_{ij}(K^+_j(r)+1)}.
\label{eq:biasedCore}
\end{equation}
where the term $1$ in the numerator and denominator has been added as it is possible that  $K^+_j (r) = 0$ if node $j$ has no links with the network's core and then the random-walk will be ill-defined.

As we want to have the best possible approximation to the maximal rate entropy $s_{\rm max}$  we define the core as the value of $r$ which maximises the value of $s(r)$, that is
\begin{equation}
\label{eq:defCore}
r_c=  \underset{r}{\mathrm{argmax}} \left( s(r) \right),
\end{equation}
where $s(r)$ is the $r$ dependent entropy
\begin{equation}
s(r)=-\sum_{i=1}^N
p^*_i(r)\sum_{j=1}^N\jumpProb(r)  \ln (\jumpProb(r))
\end{equation}
and $p^*_i(r)$ is the stationary distribution corresponding to the core--biased random jumps of Eq.~(\ref{eq:biasedCore}).
The core are the nodes ranked from $1$ to $r_c$.  The value of $s(r)$ is evaluated numerically from the core biased random jump $\jumpProb(r)$ using  Eq.~(\ref{eq:defJump}) with $f_j=K^+_j(r)+1$,  then evaluating the stationary distribution $\{p^*_i(r)\}$ (via Eq.~\ref{eq:statProb}) and from this distribution the rate entropy $s(r)$ (Eq.~(\ref{eq:entrop})) and the rank $r$ that maximises $s(r)$ (Eq.~(\ref{eq:defCore})).

Notice that the spectral--core and the core--biased random walk, even that both are formulated as a function of the density of connections between the top ranked $r$ nodes, there are different cores.

Table~\ref{tab:networks} shows the approximation to the maximal entropy using the core-biased random walk, the relative size of the core with respect to the network's size and the assortativity coefficient. The relative size of the core is not related to the assortativity of the network. 

\begin{table}
\begin{center}
\setlength\tabcolsep{4pt} 
\begin{tabular}{c c c c  }\hline
Network &  $s_c/s_{\rm max}$ &  $r_c/N$ &$\rho$  \\ 
\hline
Airports &  { 0.999} & 0.136 &  -0.267 \\ %\hline
CondMat &  { 0.945} &    0.039  & 0.157 \\ %\hline
NetSci &  { 0.914} & 0.137  & -0.081 \\ %\hline
Football &  {0.998} &  0.913  & 0.162 \\ %\hline
LesMis &  { 0.997} &  0.350  & -0.165 \\ %\hline
Random &  {0.983} & 0.449  & -0.045 \\ %\hline 
Power law &  { 0.972} & 0.227  & -0.004 \\ %\hline
Power law &  {0.970} & 0.489  & -0.245 \\ %\hline
Power law &  {0.980} &  0.126  & 0.222 \\ %\hline
Regular &  { 1.000} & 1.000 & -- \\ \hline
\end{tabular}
\end{center}
\caption{\label{tab:networks}
Ratio of the core-biased entropy against the maximal entropy ($s_c/s_{\rm max})$, relative size of the core and assortativity coefficient for some real networks.
}
\end{table}

\section*{Stationary probability for the core-biased random walk}
The stationary distribution for the core-biased random walk is evaluated using Eq.~(\ref{eq:statProb}) with $f_j=K_j^+(r)+1$ which gives
\begin{equation}
p_i^* = \frac{(K_i^+(r)+1)\sum_j ({\bf A})_{ij}(K_j^+(r)+1)}{\sum_i (K_i^+(r)+1)\sum_j ({\bf A})_{ij}(K_j^+(r)+1)},
\end{equation}
which is the probability to find the random walker in node $i$ after spending a long time visiting the network by preferring to visit nodes connected to the core.

%\bibliographystyle{abbrv}
%\bibliography{BeyondRich.bib}

\begin{thebibliography}{10}

\bibitem{alstott2014unifyingA}
J.~Alstott, P.~Panzarasa, M.~Rubinov, E.~T. Bullmore, and P.~E. V{\'e}rtes.
\newblock A unifying framework for measuring weighted rich clubs.
\newblock {\em Scientific reports}, 4:7258, 2014.

\bibitem{ball2014rich}
G.~Ball, P.~Aljabar, S.~Zebari, N.~Tusor, T.~Arichi, N.~Merchant, E.~C.
  Robinson, E.~Ogundipe, D.~Rueckert, A.~D. Edwards, et~al.
\newblock Rich-club organization of the newborn human brain.
\newblock {\em Proceedings of the National Academy of Sciences}, page
  201324118, 2014.

\bibitem{battiston2018multiplexA}
F.~Battiston, J.~Guillon, M.~Chavez, V.~Latora, and F.~D.~V. Fallani.
\newblock Multiplex core--periphery organization of the human connectome.
\newblock {\em Journal of The Royal Society Interface}, 15(146):20180514, 2018.

\bibitem{bianconi2007entropy}
G.~Bianconi.
\newblock The entropy of randomized network ensembles.
\newblock {\em EPL (Europhysics Letters)}, 81(2):28005, 2007.

\bibitem{bianconi2009entropy}
G.~Bianconi.
\newblock Entropy of network ensembles.
\newblock {\em Physical Review E}, 79(3):036114, 2009.

\bibitem{bianconi2005loops}
G.~Bianconi, G.~Caldarelli, and A.~Capocci.
\newblock Loops structure of the {I}nternet at the autonomous system level.
\newblock {\em Physical Review E}, 71(6):066116, 2005.

\bibitem{bianconi2006effect}
G.~Bianconi and M.~Marsili.
\newblock Effect of degree correlations on the loop structure of scale-free
  networks.
\newblock {\em Physical Review E}, 73(6):066127, 2006.

\bibitem{Bornholdt01}
S.~Bornholdt and H.~Ebel.
\newblock World--wide web scaling exponent from {S}imon's 1955 model.
\newblock {\em Phys. Rev. E}, 64:035104, 2001.

\bibitem{burda2009localization}
Z.~Burda, J.~Duda, J.-M. Luck, and B.~Waclaw.
\newblock Localization of the maximal entropy random walk.
\newblock {\em Physical review letters}, 102(16):160602, 2009.

\bibitem{Colizza06}
V.~Colizza, A.~Flammini, M.~A. Serrano, and A.~Vespignani.
\newblock Detecting rich-club ordering in complex networks.
\newblock {\em Nat}, 2(2):110--115, 2006.

\bibitem{collin2013impaired}
G.~Collin, R.~S. Kahn, M.~A. de~Reus, W.~Cahn, and M.~P. van~den Heuvel.
\newblock Impaired rich club connectivity in unaffected siblings of
  schizophrenia patients.
\newblock {\em Schizophrenia bulletin}, 40(2):438--448, 2013.

\bibitem{csigi2017geometric}
M.~Csigi, A.~K{\H{o}}r{\"o}si, J.~B{\'\i}r{\'o}, Z.~Heszberger, Y.~Malkov, and
  A.~Guly{\'a}s.
\newblock Geometric explanation of the rich-club phenomenon in complex
  networks.
\newblock {\em Scientific Reports}, 7(1):1730, 2017.
\newblock Create a network generating model based on geometry that creates
  rich-club.

\bibitem{d2012robustness}
G.~D'Agostino, A.~Scala, V.~Zlati{\'c}, and G.~Caldarelli.
\newblock Robustness and assortativity for diffusion-like processes in
  scale-free networks.
\newblock {\em EPL (Europhysics Letters)}, 97(6):68006, 2012.

\bibitem{10.3389/fnhum.2014.00647}
M.~A. de~Reus and M.~P. van~den Heuvel.
\newblock Simulated rich club lesioning in brain networks: a scaffold for
  communication and integration?
\newblock {\em Frontiers in Human Neuroscience}, 8:647, 2014.

\bibitem{della2013profiling}
F.~Della~Rossa, F.~Dercole, and C.~Piccardi.
\newblock Profiling core-periphery network structure by random walkers.
\newblock {\em Scientific reports}, 3, 2013.

\bibitem{dorogovtsev2010lectures}
S.~N. Dorogovtsev.
\newblock {\em Lectures on complex networks}, volume~24.
\newblock Oxford University Press Oxford, 2010.

\bibitem{estrada2011combinatorial}
E.~Estrada.
\newblock Combinatorial study of degree assortativity in networks.
\newblock {\em Physical review E}, 84(4):047101, 2011.

\bibitem{estrada2008communicability}
E.~Estrada and N.~Hatano.
\newblock Communicability in complex networks.
\newblock {\em Physical Review E}, 77(3):036111, 2008.

\bibitem{fiol2009number}
M.~A. Fiol and E.~Garriga.
\newblock Number of walks and degree powers in a graph.
\newblock {\em Discrete Mathematics}, 309(8):2613--2614, 2009.

\bibitem{fire2017rise}
M.~Fire and C.~Guestrin.
\newblock The rise and fall of network stars.
\newblock {\em arXiv preprint arXiv:1706.06690}, 2017.

\bibitem{gleiser2007become}
P.~Gleiser.
\newblock How to become a superhero.
\newblock {\em Journal of Statistical Mechanics: Theory and Experiment},
  2007(09):P09020, 2007.

\bibitem{gollo2015dwelling}
L.~L. Gollo, A.~Zalesky, R.~M. Hutchison, M.~van~den Heuvel, and M.~Breakspear.
\newblock Dwelling quietly in the rich club: brain network determinants of slow
  cortical fluctuations.
\newblock {\em Phil. Trans. R. Soc. B}, 370(1668):20140165, 2015.

\bibitem{gomez2010discrete}
S.~G{\'o}mez, A.~Arenas, J.~Borge-Holthoefer, S.~Meloni, and Y.~Moreno.
\newblock Discrete-time markov chain approach to contact-based disease
  spreading in complex networks.
\newblock {\em EPL (Europhysics Letters)}, 89(3):38009, 2010.

\bibitem{gomez2008entropy}
J.~G{\'o}mez-Garde{\~n}es and V.~Latora.
\newblock Entropy rate of diffusion processes on complex networks.
\newblock {\em Physical Review E}, 78(6):065102, 2008.

\bibitem{grayson2014structural}
D.~S. Grayson, S.~Ray, S.~Carpenter, S.~Iyer, T.~G.~C. Dias, C.~Stevens, J.~T.
  Nigg, and D.~A. Fair.
\newblock Structural and functional rich club organization of the brain in
  children and adults.
\newblock {\em PloS one}, 9(2):e88297, 2014.

\bibitem{hou2014maximum}
L.~Hou, M.~Small, and S.~Lao.
\newblock Maximum entropy networks are more controllable than preferential
  attachment networks.
\newblock {\em Physics Letters A}, 378(46):3426--3430, 2014.

\bibitem{johnson2010entropic}
S.~Johnson, J.~J. Torres, J.~Marro, and M.~A. Munoz.
\newblock Entropic origin of disassortativity in complex networks.
\newblock {\em Physical review letters}, 104(10):108702, 2010.

\bibitem{leibnitz2013maximum}
K.~Leibnitz, T.~Shimokawa, F.~Peper, and M.~Murata.
\newblock Maximum entropy based randomized routing in data-centric networks.
\newblock In {\em Network Operations and Management Symposium (APNOMS), 2013
  15th Asia-Pacific}, pages 1--6. IEEE, 2013.

\bibitem{li2011link}
R.-H. Li, J.~X. Yu, and J.~Liu.
\newblock Link prediction: the power of maximal entropy random walk.
\newblock In {\em Proceedings of the 20th ACM international conference on
  Information and knowledge management}, pages 1147--1156. ACM, 2011.

\bibitem{lin2018non}
Y.~Lin and Z.~Zhang.
\newblock Non-backtracking centrality based random walk on networks.
\newblock {\em arXiv preprint arXiv:1803.03087}, 2018.

\bibitem{lu2016drought}
X.~Lu, C.~Gray, L.~E. Brown, M.~E. Ledger, A.~M. Milner, R.~J. Mondrag{\'o}n,
  G.~Woodward, and A.~Ma.
\newblock Drought rewires the cores of food webs.
\newblock {\em Nature Climate Change}, 6:875--878, 2016.

\bibitem{ma2015rich}
A.~Ma and R.~J. Mondrag{\'o}n.
\newblock Rich-cores in networks.
\newblock {\em PloS one}, 10(3):e0119678, 2015.

\bibitem{martin2014localization}
T.~Martin, X.~Zhang, and M.~Newman.
\newblock Localization and centrality in networks.
\newblock {\em Physical Review E}, 90(5):052808, 2014.

\bibitem{maslov2002specificity}
S.~Maslov and K.~Sneppen.
\newblock Specificity and stability in topology of protein networks.
\newblock {\em Science}, 296(5569):910--913, 2002.

\bibitem{McAuley07}
J.~J. McAuley, L.~F. Costa, and T.~S. Caetano.
\newblock The rich-club phenomenon across complex network hierarchies.
\newblock {\em Appl. Phys. Lett.}, 91:084103, 2007.

\bibitem{Mondragon2014}
R.~J. Mondrag{\'o}n.
\newblock Network null-model based on maximal entropy and the rich-club.
\newblock {\em Journal of Complex Networks}, 2(3):288--298, 2014.

\bibitem{mondragon2016network}
R.~J. Mondrag{\'o}n.
\newblock Network partition via a bound of the spectral radius.
\newblock {\em Journal of Complex Networks}, page cnw029, 2016.

\bibitem{mondragon2018}
R.~J. Mondrag\'on.
\newblock Core-biased random walks in networks.
\newblock {\em Journal of Complex Networks}, page cny001, 2018.

\bibitem{Mondragon2012}
R.~J. Mondrag\'on and S.~Zhou.
\newblock Random networks with given rich-club coefficient.
\newblock {\em Eur. Phys. J. B}, 85(328), 2012.

\bibitem{Newman02}
M.~E.~J. Newman.
\newblock Assortative mixing in networks.
\newblock {\em Phys. Rev. Lett.}, 89(208701):208701, 2002.

\bibitem{nigam2016rich}
S.~Nigam, M.~Shimono, S.~Ito, F.-C. Yeh, N.~Timme, M.~Myroshnychenko, C.~C.
  Lapish, Z.~Tosi, P.~Hottowy, W.~C. Smith, et~al.
\newblock Rich-club organization in effective connectivity among cortical
  neurons.
\newblock {\em The Journal of Neuroscience}, 36(3):670--684, 2016.

\bibitem{noh2004random}
J.~D. Noh and H.~Rieger.
\newblock Random walks on complex networks.
\newblock {\em Physical review letters}, 92(11):118701, 2004.

\bibitem{ochab2013maximal}
J.~K. Ochab and Z.~Burda.
\newblock Maximal entropy random walk in community detection.
\newblock {\em The European Physical Journal Special Topics}, 216(1):73--81,
  2013.

\bibitem{Opsahl2008}
T.~Opsahl, V.~Colizza, P.~Panzarasa, and J.~J. Ramasco.
\newblock Prominence and control: The weighted rich--club effect.
\newblock {\em Phys. Rev. Lett.}, 101:168702, Oct 2008.

\bibitem{pastor2001dynamical}
R.~Pastor-Satorras, A.~V{\'a}zquez, and A.~Vespignani.
\newblock Dynamical and correlation properties of the {I}nternet.
\newblock {\em Physical review letters}, 87(25):258701, 2001.

\bibitem{rosvall2008maps}
M.~Rosvall and C.~T. Bergstrom.
\newblock Maps of random walks on complex networks reveal community structure.
\newblock {\em Proceedings of the National Academy of Sciences},
  105(4):1118--1123, 2008.

\bibitem{senden2014rich}
M.~Senden, G.~Deco, M.~A. de~Reus, R.~Goebel, and M.~P. van~den Heuvel.
\newblock Rich club organization supports a diverse set of functional network
  configurations.
\newblock {\em Neuroimage}, 96:174--182, 2014.

\bibitem{serrano2008rich}
M.~A. Serrano.
\newblock Rich-club vs rich-multipolarization phenomena in weighted networks.
\newblock {\em Physical Review E}, 78(2):026101, 2008.

\bibitem{Simon55}
H.~A. Simon.
\newblock On a class of skew distribution functions.
\newblock {\em Biometrika}, 42(3/4):425--440, 1955.

\bibitem{squartini2015breaking}
T.~Squartini, J.~de~Mol, F.~den Hollander, and D.~Garlaschelli.
\newblock Breaking of ensemble equivalence in networks.
\newblock {\em Physical review letters}, 115(26):268701, 2015.

\bibitem{squartini2011analytical}
T.~Squartini and D.~Garlaschelli.
\newblock Analytical maximum-likelihood method to detect patterns in real
  networks.
\newblock {\em New Journal of Physics}, 13(8):083001, 2011.

\bibitem{squartini2015unbiased}
T.~Squartini, R.~Mastrandrea, and D.~Garlaschelli.
\newblock Unbiased sampling of network ensembles.
\newblock {\em New Journal of Physics}, 17(2):023052, 2015.

\bibitem{Heuvel11}
M.~P. van~den Heuvel and O.~Sporns.
\newblock Rich--club organization of the human connectome.
\newblock {\em The Journal of Neuroscience}, 31(44):15775--15786, 2011.

\bibitem{van2010graph}
P.~Van~Mieghem.
\newblock {\em Graph spectra for complex networks}.
\newblock Cambridge University Press, 2010.

\bibitem{van2011n}
P.~Van~Mieghem.
\newblock The {N}-intertwined {SIS} epidemic network model.
\newblock {\em Computing}, 93(2-4):147--169, 2011.

\bibitem{van2010influence}
P.~Van~Mieghem, H.~Wang, X.~Ge, S.~Tang, and F.~Kuipers.
\newblock Influence of assortativity and degree-preserving rewiring on the
  spectra of networks.
\newblock {\em The European Physical Journal B-Condensed Matter and Complex
  Systems}, 76(4):643--652, 2010.

\bibitem{wang2003epidemic}
Y.~Wang, D.~Chakrabarti, C.~Wang, and C.~Faloutsos.
\newblock Epidemic spreading in real networks: An eigenvalue viewpoint.
\newblock In {\em 22nd International Symposium on Reliable Distributed Systems,
  2003. Proceedings.}, pages 25--34. IEEE, 2003.

\bibitem{xu2011changing}
X.~Xu, J.~Zhang, P.~Li, and M.~Small.
\newblock Changing motif distributions in complex networks by manipulating
  rich--club connections.
\newblock {\em Physica A: Statistical Mechanics and its Applications},
  390(23):4621--4626, 2011.

\bibitem{Xu10}
X.-K. Xu, J.~Zhang, and M.~Small.
\newblock Rich-club connectivity dominates assortativity and transitivity of
  complex networks.
\newblock {\em Phys. Rev. E}, 82(4):046117, Oct 2010.

\bibitem{yoon2007statistical}
S.~Yoon, S.~Lee, S.-H. Yook, and Y.~Kim.
\newblock Statistical properties of sampled networks by random walks.
\newblock {\em Physical Review E}, 75(4):046114, 2007.

\bibitem{youssef2011individual}
M.~Youssef and C.~Scoglio.
\newblock An individual-based approach to {SIR} epidemics in contact networks.
\newblock {\em Journal of theoretical biology}, 283(1):136--144, 2011.

\bibitem{yu2014maximal}
J.-G. Yu, J.~Zhao, J.~Tian, and Y.~Tan.
\newblock Maximal entropy random walk for region-based visual saliency.
\newblock {\em IEEE transactions on cybernetics}, 44(9):1661--1672, 2014.

\bibitem{zhou2008pfp}
S.~Zhou.
\newblock Why the {PFP} model reproduces the {I}nternet?
\newblock In {\em Communications, 2008. ICC'08. IEEE International Conference
  on}, pages 203--207. IEEE, 2008.

\bibitem{Zhou04}
S.~Zhou and R.~Mondrag{\'o}n.
\newblock Accurately modeling the {I}nternet topology.
\newblock {\em Physical Review E}, 70(6):066108, 2004.

\bibitem{Zlatic09}
V.~Zlatic, G.~Bianconi, A.~D{\'i}az-Guilera, D.~Garlaschelli, F.~Rao, and
  G.~Caldarelli.
\newblock On the rich-club effect in dense and weighted networks.
\newblock {\em The European Physical Journal B}, 67:271--275, 2009.

\end{thebibliography}
%\putbib
%
%\end{bibunit}

\end{document}